\documentclass[twocolumn,trackchanges]{aastex701}
\usepackage{amsmath}
\usepackage{array}
\usepackage{makecell}
\usepackage{tabularx}
\newcommand{\ha} {\mbox{H$\alpha$}\,}
\newcommand{\hb} {\mbox{H$\beta$}\,}
\newcommand{\hg} {\mbox{H$\gamma$}\,}
\newcommand{\hd} {\mbox{H$\delta$}\,}

\newcommand{\Naid}{\ion{Na}{1}\,D\,}

\newcommand{\Feii} {\ion{Fe}{2}\,}
\newcommand{\FeII}{\ion{Fe}{2}\,}
\newcommand{\ScII}{\ion{Sc}{2}\,}

\newcommand{\Scii}{\ion{Sc}{2}\,}

\newcommand{\Oi}{\ion{O}{1}\,}
\newcommand{\Baii}{\ion{Ba}{2}\,}
\newcommand{\Caii}{\ion{Ca}{2}\,}
\newcommand{\Tiii} {\ion{Ti}{2}\,}
\newcommand{\Hei} {\ion{He}{1}\,}

\newcommand{\kms}{\mbox{$\rm{km}\,s^{-1}$}}

\begin{document}

\title{SN\,2024abfl: A Low-Luminosity Type IIP Supernova at the Low-Mass End of Core Collapse}

\author[orcid=0009-0007-5354-2611]{Luhan Li}
\affiliation{Yunnan Observatories, Chinese Academy of Sciences, Kunming 650216, China}
\affiliation{International Centre of Supernovae (ICESUN), Yunnan Key Laboratory of Supernova Research, Kunming 650216, China}
\affiliation{University of Chinese Academy of Sciences, Beijing 100049, China}
\email{liluhan@ynao.ac.cn}  

\author[0000-0002-8296-2590]{Jujia Zhang}
\affiliation{Yunnan Observatories, Chinese Academy of Sciences, Kunming 650216, China}
\affiliation{International Centre of Supernovae (ICESUN), Yunnan Key Laboratory of Supernova Research, Kunming 650216, China}
\email[show]{jujia@ynao.ac.cn}

\author[0009-0005-2963-7245]{Zeyi Zhao}
\affiliation{Yunnan Observatories, Chinese Academy of Sciences, Kunming 650216, China}
\affiliation{University of Chinese Academy of Sciences, Beijing 100049, China}
\email{zhaozeyi@ynao.ac.cn}

\author[0009-0003-3758-0598]{Liping Li}
\affiliation{Yunnan Observatories, Chinese Academy of Sciences, Kunming 650216, China}
\affiliation{International Centre of Supernovae (ICESUN), Yunnan Key Laboratory of Supernova Research, Kunming 650216, China}
\email{liliping@ynao.ac.cn}

\author[0000-0002-7334-2357]{Xiaofeng Wang} 
\affiliation{Physics Department, Tsinghua University, Beijing 100084, China}
\email{wang\_xf@mail.tsinghua.edu.cn}

\author[]{Liyang Chen } 
\affiliation{Physics Department, Tsinghua University, Beijing 100084, China}
\email{chenly23@mails.tsinghua.edu.cn}

\author[]{Zeyi Wang}
\affiliation{Yunnan Observatories, Chinese Academy of Sciences, Kunming 650216, China}
\affiliation{University of Chinese Academy of Sciences, Beijing 100049, China}
\email{wangzeyi@ynao.ac.cn}

\author[]{Jingxiao Luo}
\affiliation{Yunnan Observatories, Chinese Academy of Sciences, Kunming 650216, China}
\affiliation{International Centre of Supernovae (ICESUN), Yunnan Key Laboratory of Supernova Research, Kunming 650216, China}
\affiliation{University of Chinese Academy of Sciences, Beijing 100049, China}
\email{luojingxiao@ynao.ac.cn}

\author[]{Zhengwei Liu}
\affiliation{Yunnan Observatories, Chinese Academy of Sciences, Kunming 650216, China}
\affiliation{International Centre of Supernovae (ICESUN), Yunnan Key Laboratory of Supernova Research, Kunming 650216, China}
\email{zwliu@ynao.ac.cn}

\author[]{Zhanwen Han}
\affiliation{Yunnan Observatories, Chinese Academy of Sciences, Kunming 650216, China}
\affiliation{International Centre of Supernovae (ICESUN), Yunnan Key Laboratory of Supernova Research, Kunming 650216, China}
\email{zhanwenhan@ynao.ac.cn}

\author[0000-0002-3231-1167]{Bo Wang}
\affiliation{Yunnan Observatories, Chinese Academy of Sciences, Kunming 650216, China}
\affiliation{International Centre of Supernovae (ICESUN), Yunnan Key Laboratory of Supernova Research, Kunming 650216, China}
\email[show]{wangbo@ynao.ac.cn}


\begin{abstract}

We present optical photometric and spectroscopic observations of the low-luminosity (LL) Type IIP supernova SN\,2024abfl.
The distance to its host galaxy is highly uncertain, with independent estimates of $9.5^{+2.3}_{-2.4}$ Mpc and $15.0^{+8.9}_{-1.9}$ Mpc. Even adopting the larger distance, the inferred plateau luminosity is only $\sim 10^{41}\rm erg\,s^{-1}$, placing SN 2024abfl at the extreme faint end of SNe IIP population. Its light curve exhibits a long-lasting plateau of approximately 110 days.
The spectra show exceptionally low expansion velocities, with the \FeII\, velocity of $\sim1200\,\rm km\,s^{-1}$ at 50 days after the explosion, significantly lower than the typical values of $\sim2000-5500\,\rm km\,s^{-1}$ observed in SNe IIP, placing SN\,2024abfl among the slowest-expanding LL SNe IIP.
Bolometric modeling yields a synthesized $^{56}$Ni mass of $\sim0.002-0.004\,\rm M_\odot$, though this estimate remains subject to significant uncertainty owing to the poorly constrained distance.
Considering the plateau color and duration, the magnitude drop from plateau to tail, and the progenitor luminosity, we favor a low-mass core-collapse origin for SN\,2024abfl.

\end{abstract}




\section{Introduction} \label{sect:intro}

Core-collapse (CC) supernovae (SNe) mark the explosive deaths of massive stars with initial masses $\gtrsim 8\,\rm M_\odot $, arising from the gravitational collapse of their stellar cores \citep[see e.g.,][]{2003ApJ...591..288H,2009ARA&A..47...63S}.
Type II supernovae (SNe II) are hydrogen-rich CC events that are observationally divided into Type IIP (SNe IIP), Type IIL (SNe IIL), Type IIn (SNe IIn), and Type IIb \citep[SNe IIb, e.g.,][]{1997ARA&A..35..309F}.
However, recent studies and the discovery of transitional events bridging the gap between subtypes challenge the strict subclass boundaries, revealing a continuous distribution of light-curve decline rates and suggesting that SNe II form a single and continuous family \citep[e.g.,][]{2015MNRAS.448.2608V,2017ApJ...834..118M,2021MNRAS.506.1832M,2025A&A...704A.233L}.

In the observation, SNe IIP account for about 70\% of SNe II within 40\,Mpc \citep[]{2025A&A...698A.305M}.
SNe IIP exhibit a long-lasting plateau phase of roughly three months after the explosion, followed by a rapid decline to the radioactive tail.
The durable plateau light curve is powered by the thermalization of the initial shock energy and the recombination of ionized hydrogen in the expanding envelope \citep[e.g.,][]{1993ApJ...414..712P}.

The population of SNe IIP exhibits a diverse range of intrinsic luminosities, with peak absolute $V$-band magnitudes spanning from approximately -14 to -18\,mag \citep[e.g.,][]{2014ApJ...786...67A,2025PASP..137d4203D}.
The faint tail of this luminosity distribution (plateau magnitudes between -13.5 and $\sim$ -15.5\,mag) is populated by the low-luminosity (LL) SNe II \citep[e.g.,][]{2004MNRAS.347...74P,2014MNRAS.439.2873S}.
To date, approximately 30 such objects have been presented \citep[see][]{2021A&A...655A..90Y}. 
Physically, LL SNe II are distinguished not only by their faint plateaus but also by slow expansion velocities and low late-time tail luminosities.
The observational features collectively suggest that LL SNe II are driven by low explosion energies and synthesize a small amount of radioactive $^{56}$Ni \citep[0.001-0.025\,$\rm M_\odot$,][]{2025arXiv250620068D}.
However, the exact explosion mechanism of LL SNe remains debated, with proposed channels ranging from low-mass CC SNe to electron-capture (EC) SNe.
Since LL SNe II likely originate from $\sim$8–12 $\rm M_\odot$ progenitor stars, which are just massive enough to undergo CC stage. 
This scenario is supported by pre-explosion progenitor identification \citep[e.g.,][]{Maund2005, Li2006, Mattila2008, ONeill2019, VanDyk2023}, evolutionary models of low-mass red supergiants \citep[RSG, e.g.,][]{Lisakov2017}, and simulations of SN light curves and spectra \citep[e.g.,][]{Pumo2017, Fraser2011,Jerkstrand2018}.

The EC SN channel is particularly relevant for LL SNe.
Unlike standard CC explosions of RSGs, ECSNe triggered by electron capture onto ONeMg cores in super-asymptotic giant branch (SAGB) stars are characterized by a steep density gradient outside the core, which ensures a successful explosion even with low energy \citep[e.g.,][]{Miyaji+1980,Nomoto1982Natur.299..803N,Nomoto1987ApJ,2009ApJ...705L.138P,Doherty2015MNRAS.446.2599D,Zha2019ApJ...886...22Z,Tauris2023pbse.book.....T,2026RAA....26c2001W}.
\citet{2009ApJ...705L.138P} explored the parameter space and explosion parameters for the formation of ECSNe from SAGB stars and discussed the potential connection between this channel and SN\,2008ha and SN\,2008S.
Some simulations predict low explosion energies ($\sim 10^{50}$\,erg) and small $^{56}\text{Ni}$ yields ($\rm \sim 0.002-0.015 \,M_\odot$), which are an order of magnitude lower than those of typical CCSNe  \citep[e.g.,][]{Timmes1996ApJ...457..834T,Kitaura2006A&A...450..345K, Wanajo2009ApJ...695..208W, Fryer2012ApJ...749...91F}. 
Despite their low explosion energies, ECSNe from SAGB progenitors may be observed as relatively faint SN II (IIP or IIL, depending on the mass of the H-rich envelope) or stripped-envelope SNe \citep{2009ApJ...705L.138P}.
Therefore, ECSNe are considered a possible formation channel for LL SNe IIP.

RSGs have been directly identified and characterized in pre-explosion images for several SNe II.
To date, approximately 20 progenitors of SNe IIP have been detected through such observations \citep[see][]{2025Galax..13...33V}.
For instance, \citet{2005MNRAS.364L..33M} identified the progenitor of SN\,2005cs as a mid-K to late-M type RSG using Hubble Space Telescope (HST) imaging.
By mapping its spectral energy distribution onto a theoretical H-R diagram, they constrained the initial mass to $7$--$12\,\rm M_\odot$.
Similarly, \citet{2024ApJ...969L..15X} derived an initial mass of $12$--$14\,\rm M_\odot$ for the progenitor of SN\,2024ggi using comparable methods. 
\citet{2025ApJ...982L..55L} recently identified the progenitor of SN\,2024abfl in archival HST data, suggesting an RSG with an initial mass of $9$-$12\,\rm M_\odot$ (adopting a distance of $15.6^{+6.1}_{-3.0}\,$Mpc).

In this paper, we present photometric and spectroscopic observations of SN\,2024abfl, a peculiar and LL SN IIP characterized by a durable plateau and low envelope velocities. 
Taking into account the different distance estimates derived from various methods, we suggest that SN\,2024abfl possesses a low absolute magnitude and a small $^{56}$Ni mass, placing it at the faint end of the LLSN IIP population. 
Relevant data reduction techniques are detailed in Section~\ref{sect:obs}.
In Section~\ref{sect:basic_inf}, we report the distance, extinction, and reddening associated with the host galaxy of SN\,2024abfl. 
The photometric and spectroscopic analyses are presented in Sections~\ref{sect:photometry} and~\ref{sect:spectroscopy}, respectively. 
Finally, the discussions are presented in Section~\ref{sect:discussion}, and the conclusions are provided in Section~\ref{sect:conclusion}.
Additionally, we present the supplementary figure in Appendix~\ref{sect:supfigure}, and the relevant data tables in Appendix~\ref{sect:suptable}.

\section{Observations and data reduction} \label{sect:obs}

\subsection{Photometric data}
\begin{figure}
   \centering
   \includegraphics[width = \linewidth]{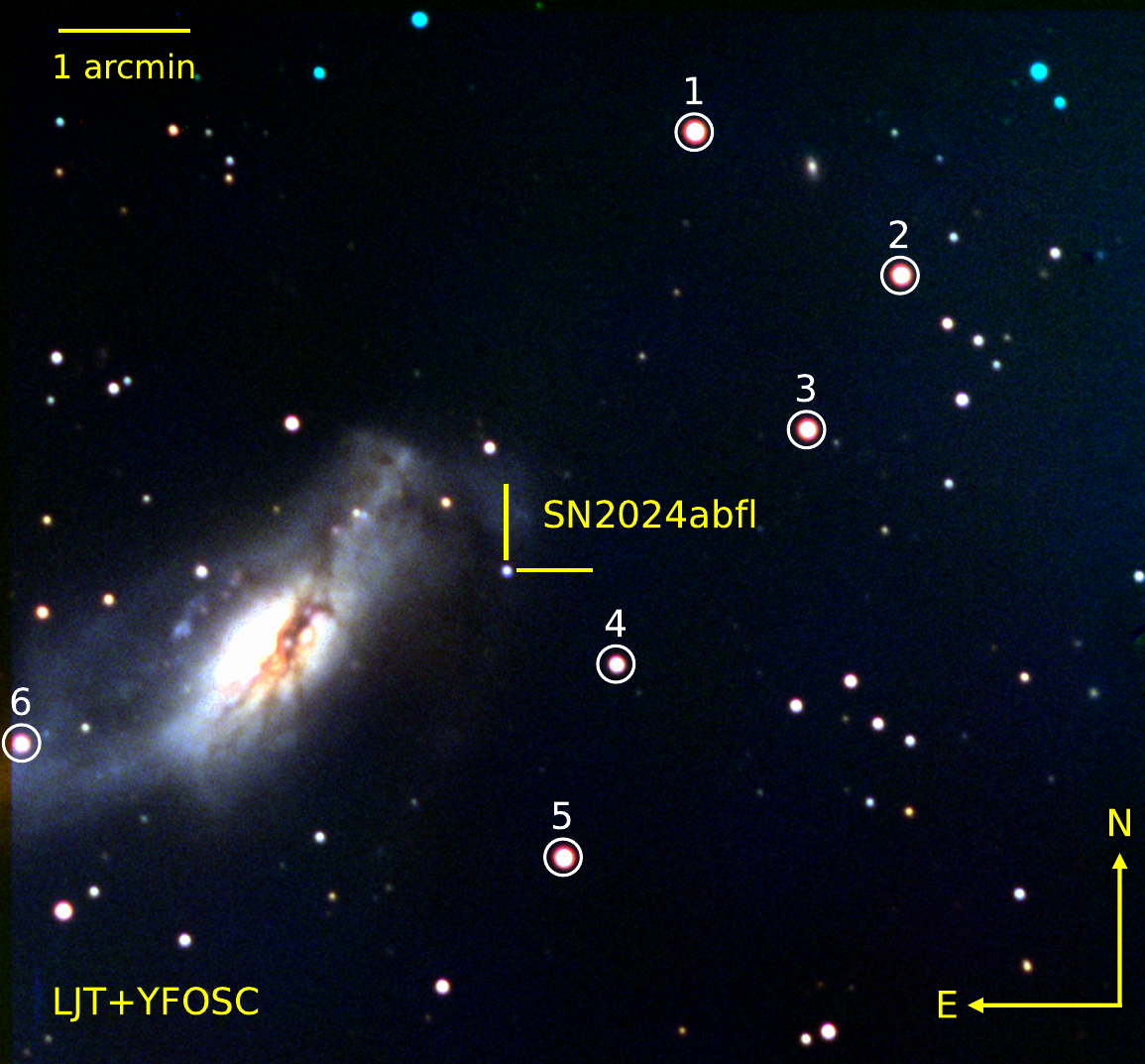}
      \caption{Image of the location of SN\,2024abfl in NGC\,2146, obtained with the LJT in the $BVr$ filters by combining data from multiple epochs.
      The local reference stars listed in Table~\ref{table:reference_star} are marked with a nearby number.
      The orientation and scale of the images are reported.}
     \label{Fig:location}
\end{figure}

Optical photometry of SN\,2024abfl in the \textit{uBVgriz} bands was obtained by the Lijiang 2.4\,m telescope equipped with the Yunnan Faint Object Spectrograph and Camera \citep[LJT+YFOSC;][]{2015RAA....15..918F,2019RAA....19..149W} from modified Julian Dates (MJD) 60630.7 to 60775.5.
The photometric data of LJT are listed in Table~\ref{table:photdata}.
Aperture photometry of SN\,2024abfl and the comparison stars was performed after template subtraction using standard \texttt{IRAF}\footnote{\texttt{IRAF}, the Image Reduction and Analysis Facility, is distributed by the National Optical Astronomy Observatory, which is operated by the Association of Universities for Research in Astronomy (AURA), Inc. under cooperative agreement with the U.S. National Science Foundation (NSF).} routines. 
The Zwicky Transient Facility (ZTF) started monitoring SN\,2024abfl (named ZTF24abtczty) on 2024-11-05 in \textit{g} band.
For a data complement, we download the public data\footnote{\url{https://alerce.online/object/ZTF24abtczty}} of ZTF from the start to 2025-05-09 \citep{2019PASP..131a8003M}. 
Additionally, we also collected photometric data of \textit{c} and \textit{o} bands from the Asteroid Terrestrial-impact Last Alert System (ATLAS) sky surveys for SN\,2024abfl \citep{2018PASP..130f4505T,2020PASP..132h5002S}.
The forced photometry light curves were directly obtained from the ATLAS data-release server \footnote{\url{https://fallingstar-data.com/forcedphot/}} \citep{Shingles2021TNSAN...7....1S}. 
To improve the data quality, we employed the public stacking tool\footnote{\url{https://github.com/thespacedoctor/plot-results-from-atlas-force-photometry-service}} to combine the single-filter measurements into daily stacks, which performs sigma-clipping to remove outliers.
Moreover, the UV photometry data obtained with UVOT on the Neil Gehrels Swift Observatory \citep{2004ApJ...611.1005G,2005SSRv..120...95R} are also included here.
We reduced the data using the HEASoft\footnote{\url{https://heasarc.gsfc.nasa.gov/docs/software/lheasoft/}} software package. 
Photometry was performed using the uvotsource task with a standard 5-arcsec circular aperture, and the background was estimated from a nearby source-free region.
Magnitudes are reported for detections with signal-to-noise ratio (S/N) $>3\sigma$, while $3\sigma$ upper limits are given for non-detections.

\subsection{Spectroscopic data}

The spectra of SN\,2024abfl were obtained with the LJT \citep{Wang-2019RAA} and Xinglong 2.16m telescope \citep[XLT,][]{2016PASP..128j5004Z}.
Information for all spectroscopic observations is listed in Table~\ref{table:specinfo}.
All spectra obtained by LJT and XLT were reduced using standard IRAF routines. 
The spectra were corrected for the local atmospheric extinction and calibrated with spectrophotometric standard stars observed at a similar airmass on the same night. 
The telluric lines were also removed.
Additionally, a public early-time spectrum obtained from Gemini-N/GMOS on 2024-11-16 was retrieved from the Weizmann Interactive Supernova Data Repository \citep[WISeREP;][]{2012PASP..124..668Y} to improve the spectral evolution coverage.

\section{Basic target information} \label{sect:basic_inf}

\begin{table}
\centering
\caption{Summary of the basic properties of SN\,2024abfl}
\label{tab:properties}
\renewcommand{\arraystretch}{1.1}
\begin{tabular}{@{}lr@{}} 
\hline\hline
RA (J2000) & 06:18:01.140\\
DEC (J2000) & +78:22:01.52\\
Host Galaxy & NGC\,2146\\
Redshift $z$ & $0.002979$\\
$E(B - V)_{\mathrm{Gal}}$$^{1}$ & $0.085 \, \mathrm{mag}$\\
$E(B - V)_{\mathrm{host}}$ & $0.105 \, \mathrm{mag}$\\
Estimated Explosion (MJD)$^{\mathrm{2}}$ & $60623.16^{+0.14}_{-0.07}$\\
$\rm High\ distance\ (Mpc)$ & $15.0^{+8.9}_{-1.9}$ \\
$\rm Low\ distance$\ (Mpc) & $9.5^{+2.3}_{-2.4}$ \\
\hline
\end{tabular}
\vspace{0.1mm}
\begin{flushleft}
Notes:
1. Retrieved from NASA/IPAC NED \citep{Schlafly2011ApJ...737..103S}. \\
2. Estimated from the expanding fireball model.
\end{flushleft}
\end{table}

\subsection{Location}
SN\,2024abfl was discovered by the \citet{2024TNSTR4506....1I} on 2024-11-15.58 (epoch corresponding to MJD = 60629.58, UT dates are used throughout this paper), at a clear filter of $ 17.5\, \mathrm{mag}$.
It was quickly classified as a young SN II by the Himalayan Chandra telescope \citep[HCT;][]{2024TNSCR4515....1D}.
Its J2000 coordinates are $\rm RA\ 06^{h}18^{m}01.140^{s}$, $\rm Dec.\ +78^{\circ}22^{\prime}01.52^{\prime\prime}$, placing it $36.24^{\prime\prime}$ north and $9.14^{\prime}$ west of the core of the nearby SB(s)ab galaxy NGC\,2146.
The location of SN\,2024abfl within the host galaxy is illustrated in Figure~\ref{Fig:location}, and the basic properties of SN\,2024abfl are shown in Table~\ref{tab:properties}.
Interestingly, in the same host galaxy, the J2000 coordinates of post-explosion SN\,2018zd are $\rm RA\ 06^{h}18^{m}03.190^{s}$, $\rm Dec.\ +78^{\circ}22^{\prime}01.16^{\prime\prime}$, which located only $\sim 5.1^{\prime\prime}$ away from SN\,2024abfl.
This close angular distance makes the analysis of SN\,2018zd and SN\,2024abfl critical for understanding the progenitor environment and explosion mechanism in this peculiar starburst galaxy.

\subsection{Extinction}

Regarding the interstellar reddening, we adopt $E(B-V)_{\rm MW} = 0.085$\,mag for the Galactic reddening contribution \citep{Schlafly2011ApJ...737..103S}, retrieved via the NASA/IPAC Extragalactic Database (NED), assuming a reddening law with $R_V = 3.1$ \citep{Cardelli1989ApJ...345..245C}.
Two \Naid absorption features are detected in the early-time spectra of SN\,2024abfl from Gemini-N/GMOS obtained on 2024-11-16, as shown in Figure~\ref{Fig:NaID}.
The equivalent widths (EW) of the Milky Way and host-galaxy components are approximately 1.01\,\AA\  and 0.74\,\AA, respectively.
The features were fitted using a composite model consisting of a linear continuum and two Gaussian absorption profiles.
It is worth noting that the measured EWs have relatively large uncertainties due to the limited S/N ratio of the spectrum.

Several studies have reported empirical correlations between the Na I D absorption and interstellar extinction or reddening \citep[e.g.,][]{2003fthp.conf..200T,2012MNRAS.426.1465P,2018A&A...609A.135S}.
\citet{2003fthp.conf..200T} identified two branches in the relation between reddening and \Naid EW (their Figure~3).
Using the lower branch relation, 
\begin{equation}
E(B-V)=0.16 \times \mathrm{EW}-0.01.
\end{equation}
\citet{2012MNRAS.426.1465P} also proposed an empirical relation of EW and extinction
\begin{equation}
\log _{10}(E(B-V))=1.17 \times \mathrm{EW}-1.85 \pm 0.08.
\end{equation}
The two relations give a similar host-galaxy extinction of $E(B-V)_{\rm host}=0.10\,$mag.
Therefore, we adopted a mean host-galaxy extinction, yielding a total extinction of $E(B-V)_{\rm tol}=0.19\,$mag, which is used in the subsequent analysis.


\subsection{Explosion epoch}

The first multi-band photometric observation with LJT was obtained at MJD = 60630.7.
Because of the limited coverage during the rising phase of SN\,2024abfl, we collected early-time ATLAS $o$-band data (including both detections and upper limits) together with ZTF $g$-band forced photometry.
To better constrain the explosion epoch of SN\,2024abfl, we performed a simple expanding fireball model, $F(t)=F_1 \times\left(t-t_0\right)^2$, to fit the ATLAS $o$-band detections and limits at $\rm t < 10\,d$ except for the earliest ZTF detection.
The fitting results are presented in Figure~\ref{Fig:explosion_epoch}.
The explosion epoch derived from the expanding fireball model is $\rm MJD = 60623.16^{+0.14}_{-0.07}$, which is adopted in the following analysis.

\subsection{Distance}

\begin{table*}
\centering
\small
\setlength{\tabcolsep}{3pt}
\caption{Distance for NGC\,2146.}
\label{table:distance}
\begin{tabularx}{\textwidth}{c c c c} 
\hline
\hline
Distance (Mpc) & Method & Reference & Note \\
\hline
16.7$\pm$1.2 & Kinematic distance & NED & Recessional velocity is 1219$\,\rm km\,s^{-1}$\\
10.3-39.7 & Tully-Fisher distance & NED & Strong tidal interactions with a LSB companion galaxy\\
14.6$\pm$2.3 & Globular clusters distance & \citet{2021arXiv210912943C} & Globular cluster radius is 2.66 pc\\
$>$7 & TRGB distance & \citet{2021arXiv210912943C} &  \\
6.5$\pm$0.7 & 2018zd EPM & \citet{2021NatAs...5..903H} & Early spectra are dominated by CSM interaction\\
9.6$\pm$1.0 & 2018zd SCM & \citet{2021NatAs...5..903H} & \\
8.73$\pm$2.20 & 2024abfl SCM & This work & \\
$9.95^{+5.13}_{-4.35}$ & 2024abfl EPM & This work &  \\
\hline
\end{tabularx}
\end{table*}

In the same host galaxy, SN\,2018zd occurred close to SN\,2024abfl and was proposed as one of the first candidates for an electron-capture supernova \citep[ECSN;][]{2021NatAs...5..903H}. 
The distance to NGC\,2146 is crucial for constraining the progenitor mass, absolute magnitude, and explosion energy of SN\,2018zd and SN\,2024abfl, and thus directly affects the assessment of how likely SNe are to be classified as an ECSN.
In fact, the distance to ther host galaxt NGC\,2146 has attracted considerable discussion and remains controversial \citep[e.g.,][]{2021arXiv210912943C,2020MNRAS.498...84Z,2021NatAs...5..903H}.

Adopting a standard cosmology \citep[$\rm H_0=73\pm5\,kms^{-1} \, Mpc^{-1}$, $\Omega_M=0.27$ and $\Omega_{\Lambda}=0.73$;][]{Spergel2007ApJS..170..377S} and corrected for the Virgo Cluster, the Great Attractor, and the Shapley supercluster influence, we obtained a recessional velocity of $1219\pm16\, \kms$ and Hubble flow distance $d = 16.7\pm1.2\, \mathrm{Mpc}$ for NGC\,2146 \footnote{\url{https://ned.ipac.caltech.edu}} \citep{2000ApJ...529..786M}.
However, typical galactic peculiar velocities are $\sim 300\, \kms$ \citep[e.g.,][]{2009ApJS..185...32K,2011ApJ...741...67D}, and the quoted uncertainty does not account for the additional scatter introduced by such random motions.

\citet{2021NatAs...5..903H} obtained a distance of $9.6\pm1.0\,\mathrm{Mpc}$ for SN\,2018zd using the SCM.
They also reported an EPM distance of $6.5\pm0.7\,\mathrm{Mpc}$; however, the early emission of SN\,2018zd is strongly influenced by CSM interaction, making the EPM estimate unreliable.

NED provided a broad distance range of 10.3-39.7\,Mpc based on the Tully-Fisher method, which may be affected by strong tidal interaction between NGC\,2146 and its low-surface-brightness companion galaxy \citep{2001A&A...365..360T}.
\citet{1997A&A...326..915T} argued that the CO-line Tully–Fisher relation provides a more reliable distance estimate for interacting systems and obtained a distance of 23.1\,Mpc\footnote{We note that this value has been cited as 21.3\,Mpc in some literature, but the original value in \citet{1997A&A...326..915T} is 23.1\,Mpc. This correction impacts the joint probability density function of the final distance estimate.}.

In addition, \citet{2021arXiv210912943C} derived a distance of $14.6\pm2.3$\,Mpc based on globular cluster distance and suggested that the tip of the red giant branch (TRGB) distance is larger than 7\,Mpc.
They also examined the three independent distance estimates (kinematic, Tully-Fisher, and globular cluster radii) and inferred a distance of $15.6^{+6.1}_{-3.0}\,$Mpc based on the joint probability distribution.

Thanks to the fortunate occurrence of SN\,2024abfl, additional distance-determination methods for SNe IIP can be applied, allowing us to obtain a extra distance estimate.
We therefore derive the distance to SN\,2024abfl using the expanding photosphere method \citep[EPM; e.g.,][]{1974ApJ...193...27K} and the standard candle method \citep[SCM; e.g.,][]{2002ApJ...566L..63H}, adopting the average value as the final distance.

Based on the empirically observed luminosity-velocity correlation, which indicates that more luminous SNe II have higher photospheric expansion velocities, the distance to an SN II can be inferred from its apparent magnitude and envelope velocity \citep[e.g.,][]{2002ApJ...566L..63H}.
The apparent $V$-band magnitude and $v_{\text{FeII\,5169}}$ at 50\,d of SN\,2024abfl were interpolated from the +49.6\,d photometric data (transformed from $gr$ bands) and the spectrum.
This calculation assumes that the ejecta velocity and $V$-band photometry remained essentially constant within the small interval between +49.6 days and +50 days.
Using the apparent V-band magnitude ($m^{50}_{V}= 17.20\pm0.04 \,$mag) and \FeII\ velocity ($v_{\text{FeII\,5169}}= 1224.0 \pm 236.2\, \kms$ ) at 50\,d after the explosion, the SCM distance of SN\,2024abfl is estimated to be $ 8.73\pm2.20 \,$Mpc.

The EPM, developed by \citet{1974ApJ...193...27K}, is based on a simplified model in which the SN is a sharply-defined, spherically-symmetric, expanding photosphere.
While the radiation emanating from this photosphere is initially treated as blackbody emission, it is diluted by a factor $\xi_{\nu}$. The dilution factor effectively parameterizes and accounts for all spectral deviations from a perfect blackbody, including features such as lines and limb-darkening.
For simplicity in applying the EPM, it is assumed that the dilution factor depends solely on the color temperature.
Given the divergent formulas for the dilution factor presented in \citet{1996ApJ...466..911E} (E96 from now on), \citet{2005A&A...439..671D} (D05), and \citet{2019A&A...621A..29V} (V19), we utilized all three formulations to conduct the EPM fitting.
The expansion velocity $v_{\rm ph}$ is inferred based on the relation between photospheric velocities and \hb velocity from \citet{2012MNRAS.419.2783T}.
Due to the EPM's requirement for early-time data, we utilized only the parameters obtained within 10 days of the explosion for the fitting procedure.
The EPM results and quantities obtained using these three models are presented in Figure~\ref{fig:EPM_fitting} and Table~\ref{table:EPM quantities}.
The EPM provides a mean distance of $9.95^{+5.13}_{-4.35}\,\mathrm{Mpc} $ and the explosion epoch of $60625.7^{+2.6}_{-2.6}$.
The distance uncertainty is relatively large, mainly due to the dispersion of the photospheric velocity transform and the measurement error of the \hb\ absorption velocity from the spectra.
The EPM distance of SN\,2024abfl is consistent with both the SCM distance of SN\,2024abfl ($8.73\pm2.20$ Mpc) and the SCM distance of SN\,2018zd \citep[$9.6\pm1.0$ Mpc; ][]{2021NatAs...5..903H}.
The EPM explosion epoch is slightly later than the epoch estimated from the expanding fireball model ($60623.16$), but it is consistent within the 1$\sigma$ uncertainty of the EPM fit.

We collected these available distance measurements to NGC\,2146 in Table~\ref{table:distance}.
In principle, we could combine their individual probability distributions to construct a single joint probability density function (PDF) for the distance.
However, the distances inferred from the SCM and EPM methods are significantly smaller than those obtained from the other methods.
Considering this discrepancy, we constructed two separate PDFs: one based on the SNe IIP methods (SCM and EPM) and another based on the remaining distance indicators.
We show the combined PDFs for these two groups in Figure~\ref{fig:PDFs}.
The peaks of the resulting distributions yield distances of $9.5^{+2.3}_{-2.4}\,$Mpc and $15.0^{+8.9}_{-1.9}\,$Mpc, respectively,  where the uncertainties correspond to the 16th and 84th percentiles of each PDF.
Hence, both distances are adopted throughout this paper.
It is worth noting that a purely methodological comparison of distance estimates cannot determine which measurement is more reliable; instead, the following discussion based on physical considerations helps to identify the more credible value.

\section{Photometry} \label{sect:photometry}

\subsection{Apparent light curves} \label{sect:Apperent_LC}

\begin{figure*} 
   \centering
   \includegraphics[width = 0.8\textwidth]{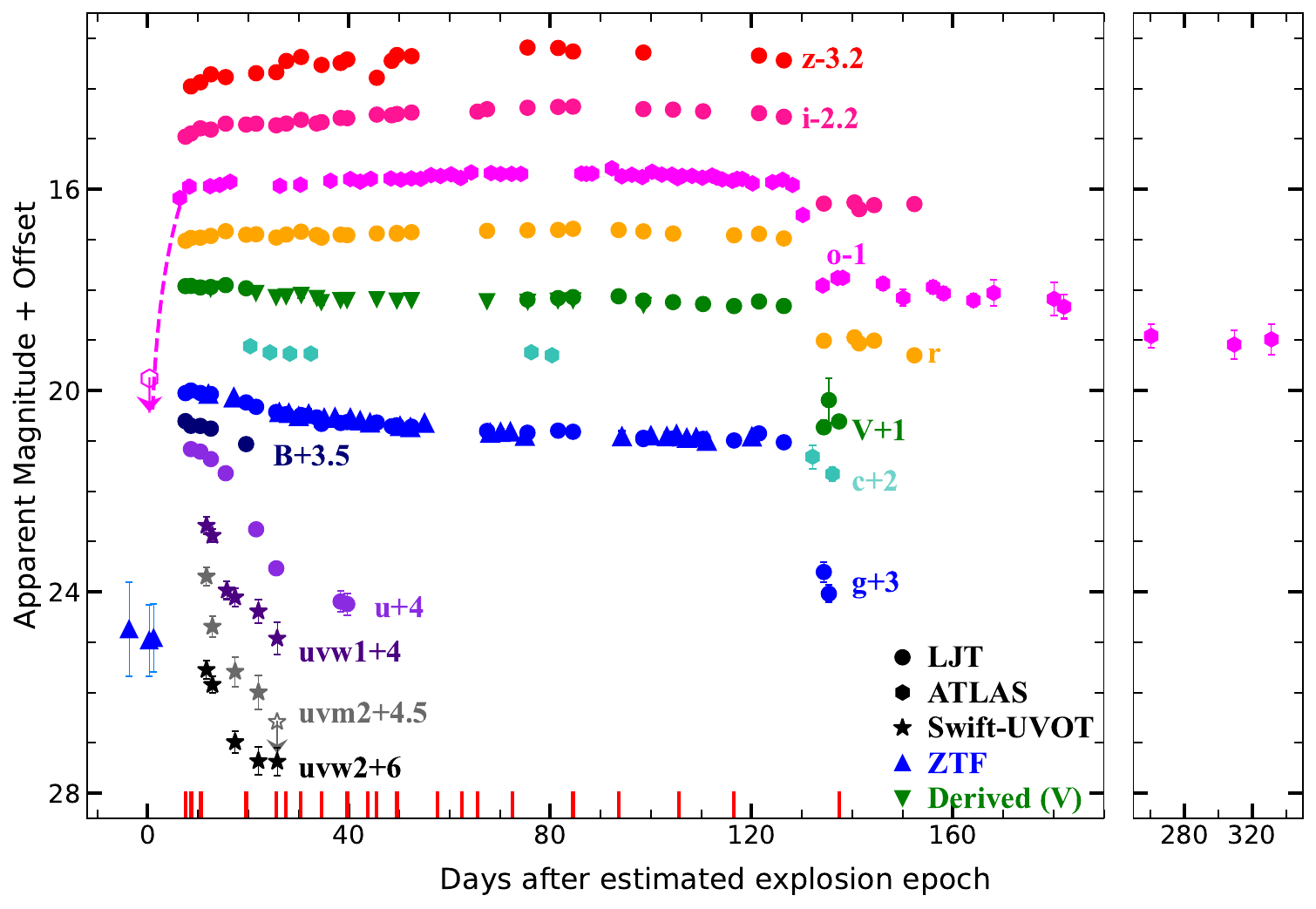}
   \caption{Multi-band apparent light curves of SN\,2024abfl obtained with the LJT, ATLAS, ZTF, and Swift/UVOT.
   All measurements are shifted vertically for better visibility.
   The adopted explosion epoch is $\rm MJD = 60623.16$.
   ATLAS data have been combined into daily stacks. 
   The best-fit expanding fireball model for $o$ band is also plotted here. 
   The direct $V$-band data with LJT is lacking around 20-80 days, so we employed the transformation equations for stars from \cite{2005AJ....130..873J} to improve the coverage of the $V$-band light curve (green inverted triangles).
   The epochs of our spectra are marked with vertical solid red lines at the bottom.
   In most cases, the photometric uncertainties are smaller than the symbol sizes.}
    \label{Fig:LC}
\end{figure*}  

We conducted continuous photometric observations of SN\,2024abfl for about five months following its discovery.
The multi-band optical light curves of SN\,2024abfl are shown in Figure~\ref{Fig:LC}, where the right panel displays the late-time $o$-band photometry around 300\,d after the explosion.
To enhance the S/N ratio and the quality of the light curve, the ATLAS $oc$-bands data were stacked. 
Due to the lack of mid-time $V$-band data, we employed the transformation equations for stars from \cite{2005AJ....130..873J} to extend the middle-time coverage of the $V$-band light curve (light green pentagon).
The $V$‑band data obtained through transformation are in good agreement with the actual $V$‑band observations, with only minor differences ($< 0.1\,$mag).

Owing to the poorly defined peak in the multi-band light curves and the incomplete observational coverage during the rising phase, the multi-band light curves do not yield a well-constrained peak magnitude.
We therefore estimated a rough $V$-band peak apparent magnitude of $16.91\pm0.01\,$mag with a rise time of $\sim15.5\,$d.
It should be noted that this peak corresponds only to the brightest observed data point, and the true uncertainty is larger than here.

Within the $gVrioz$-band light curve, a durable and flat plateau phase of approximately 110 days is identified following the definition method of \citet{2025arXiv250620068D}.
In the ZTF LL SNe IIP samples, \citet{2025arXiv250620068D} reported a $r$-band plateau duration of 64-108 days, while SN\,2024abfl located near the upper end of this range.
For quantitative analysis, we adopted linear regression to determine the decline rates in various phases and bands, which are summarized in Table~\ref{tab:decline_rate}.
The $V$-band decline rates of SN\,2024abfl during the plateau phase are relatively small, consistent with the median plateau decline rates of the LLSNe sample \citep[$\rm 0.1^{+0.1}_{-0.2}\,mag\,(100\, d)^{-1}$; ][]{2025arXiv250620068D}.
The $roiz$-band light curves exhibit a relatively flat plateau with a mild broad hump between $\sim$20 and 120 days.
This mild brightening between $\sim$20-80\,d may be primarily attributed to the gradual cooling of the photosphere, which shifts the spectral energy distribution toward longer wavelengths.
This feature is a common feature in many SNe IIP samples, as seen in events like SN\,2005cs \citep{2009MNRAS.394.2266P} and SN\,2006bc \citep{2024A&A...692A..95A}.

After the plateau phase, a distinct drop could is observed in $gVroi$ bands of SN\,2024abfl.
The plateau-tail Drop in the $Vro$ bands of SN\,2024abfl are about 2.4, 2.1, and 2.0\,mag, respectively.
For comparison, the median $r$-band plateau-tail drop for ZTF LLSNe is $1.33^{+0.72}_{-1.25}$ \citep{2025arXiv250620068D}, while SN\,2024abfl lies near the upper boundary of this distribution.

\begin{table}
    \centering
    \caption{Decline rates (in $\rm mag\, /100\,d$) of the multi-band light curves of SN\,2024abfl.}
    \renewcommand{\arraystretch}{1.0}
    \setlength{\tabcolsep}{5pt}
    \begin{tabular}{cccc}
        \hline \hline
        Band & $t_{\rm start} (\mathrm{d}) $ & $t_{\rm stop} (\mathrm{d})$ & Slope ($\rm mag\, /100\,d$)\\
        \hline
        u & 8 & 40 &11.93$\pm$0.40 \\
        B & 8 & 20 &3.88$\pm$0.17 \\
        g & 50 & 120 &0.43$\pm$0.02 \\
        c & 20 & 80 &0.12$\pm$0.03 \\
        V & 20 & 120 &0.18$\pm$0.02 \\
        \hline
        r & 20 & 85 & $-0.21\pm0.01 $\\
        r & 85 & 120 & $0.39\pm0.06 $\\
        r & 130 & 150 &1.12$\pm$0.60\\
        \hline
        o & 20 & 85 &-0.48$\pm$0.03\\
        o & 85 & 120 &0.36$\pm$0.03\\
        o & 135 & 330 &0.77$\pm$0.10\\
        \hline
        i & 20 & 85 &-0.60$\pm$0.01\\
        i & 85 & 120 &0.37$\pm$0.08\\
        i & 130 & 150 &0.26$\pm$0.41\\
        \hline
        z & 20 & 85 &-0.73$\pm$0.02\\
        z & 85 & 125 &0.22$\pm$0.05\\  
        \hline
    \end{tabular}
    \label{tab:decline_rate}
\end{table}

\subsection{Absolute light curve and comparisons}

\begin{figure}
    \centering
    \includegraphics[width=0.98\linewidth]{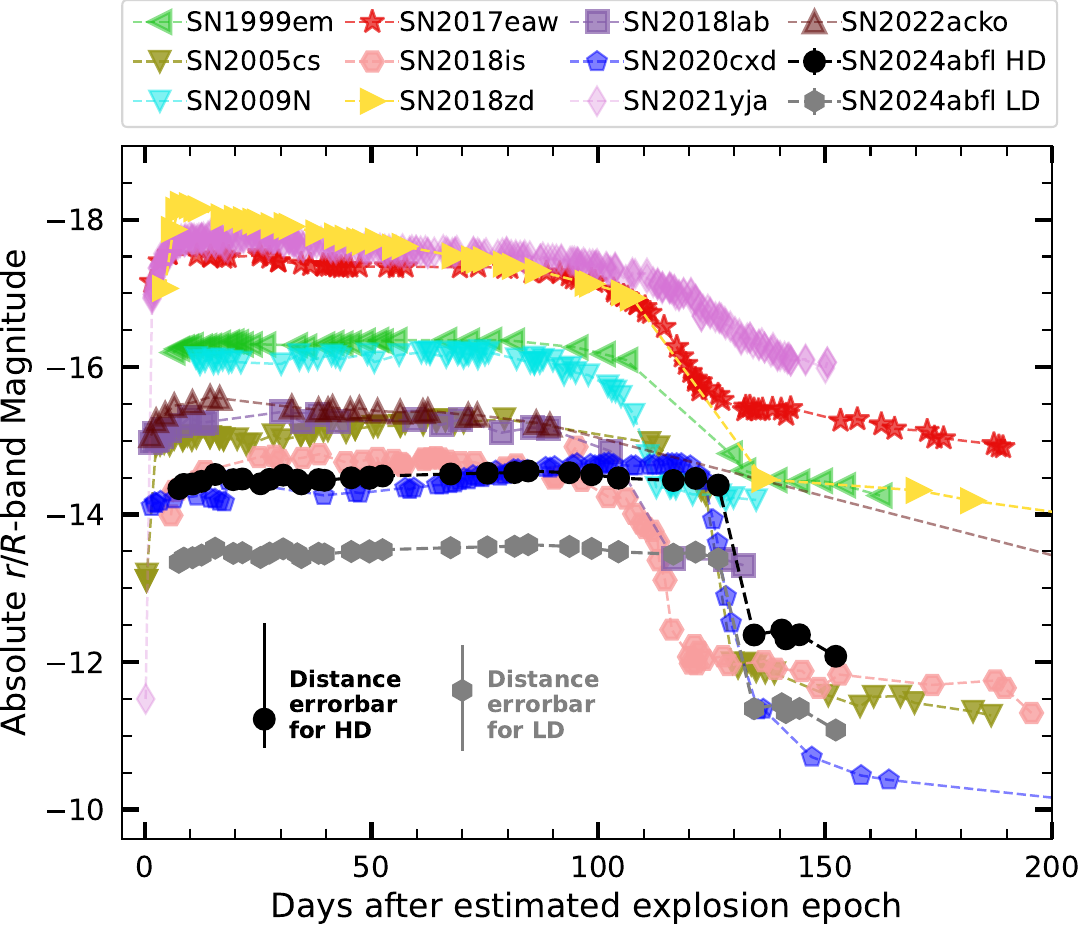}
    \caption{Comparison of the absolute $r/R$-band light curve of SN\,2024abfl with those well-observed SNe II.
    Two distances derived from different methods are considered for SN\,2024abfl. 
     Due to the large distance uncertainty of SN\,2024abfl, we show only the magnitude error bar associated with the distance in the lower left of the figure. }
    \label{fig:Abs_r}
\end{figure} 

In Figure~\ref{fig:Abs_r}, we compared the absolute $r/R$-band light curves of a subset of SNe IIP, whose basic properties are summarised in Table~\ref{tab:SNe_II info}.
We selected several representative SNe IIP for comparison, and the basic properties of all comparison objects are summarized in Table~\ref{tab:SNe_II info}.
These SNe were chosen because they have comprehensive photometric and spectroscopic observations.

In Figure~\ref{fig:Abs_r}, the $r$-band peak absolute magnitude of SN\,2024abfl is $M_{\mathrm{r}}=-14.63$ and  $M_{\mathrm{r}}=-13.63$ for the higher and lower distance estimates, respectively.
The uncertainty in magnitude due to the distance error is shown in the lower-left corner of the figure.
The luminosity of SN\,2024abfl based on the higher distance is slightly lower than the typical LLSNe, such as SN\,2005cs \citep[$M_{\mathrm{R}}=-15.5\pm0.2\,\mathrm{mag}; $][]{2006MNRAS.370.1752P} and SN\,2018is \citep[$M_{\mathrm{r}}=-14.8\pm0.2\,\mathrm{mag}; $][]{2025A&A...694A.260D}.
It is worth noting that the absolute light curve of SN\,2024abfl resembles that of the underluminous SN\,2020cxd, with both objects exhibiting comparable plateau widths and magnitudes.
However, if the lower distance is adopted, SN\,2024abfl becomes significantly fainter than these typical SNe IIP and LLSNe.
\citet{2015ApJ...799..208S} constructed the absolute magnitude distribution of SNe IIP based on Pan-STARRS and reported a mean $r$-band absolute magnitude of $-18.1^{+0.63}_{-1.18}\,$mag.
In contrast, SN\,2024abfl is significantly fainter, regardless of whether a higher or lower distance is adopted.

Given that SN\,2018zd and SN\,2024abfl reside in the same host galaxy, distance-related uncertainties are eliminated. 
Assuming negligible the difference of host galaxy extinction, this allows for a direct comparison of their apparent magnitudes.
SN\,2018zd exhibits a significantly brighter $r$-band peak luminosity, with a peak apparent magnitude of 13.6\,mag compared to $\sim$16.8\,mag for SN\,2024abfl. 
Furthermore, SN\,2018zd displays a more rapid decline rate during the plateau phase and a more pronounced magnitude drop from the peak to the end of the plateau \citep{2020MNRAS.498...84Z}.
Although the two SNe occur in the same host galaxy and are located in close proximity, they exhibit substantial differences in their light curve evolution.


\subsection{Color evolution}

\begin{figure}
   \centering
   \includegraphics[width = 0.98\linewidth]{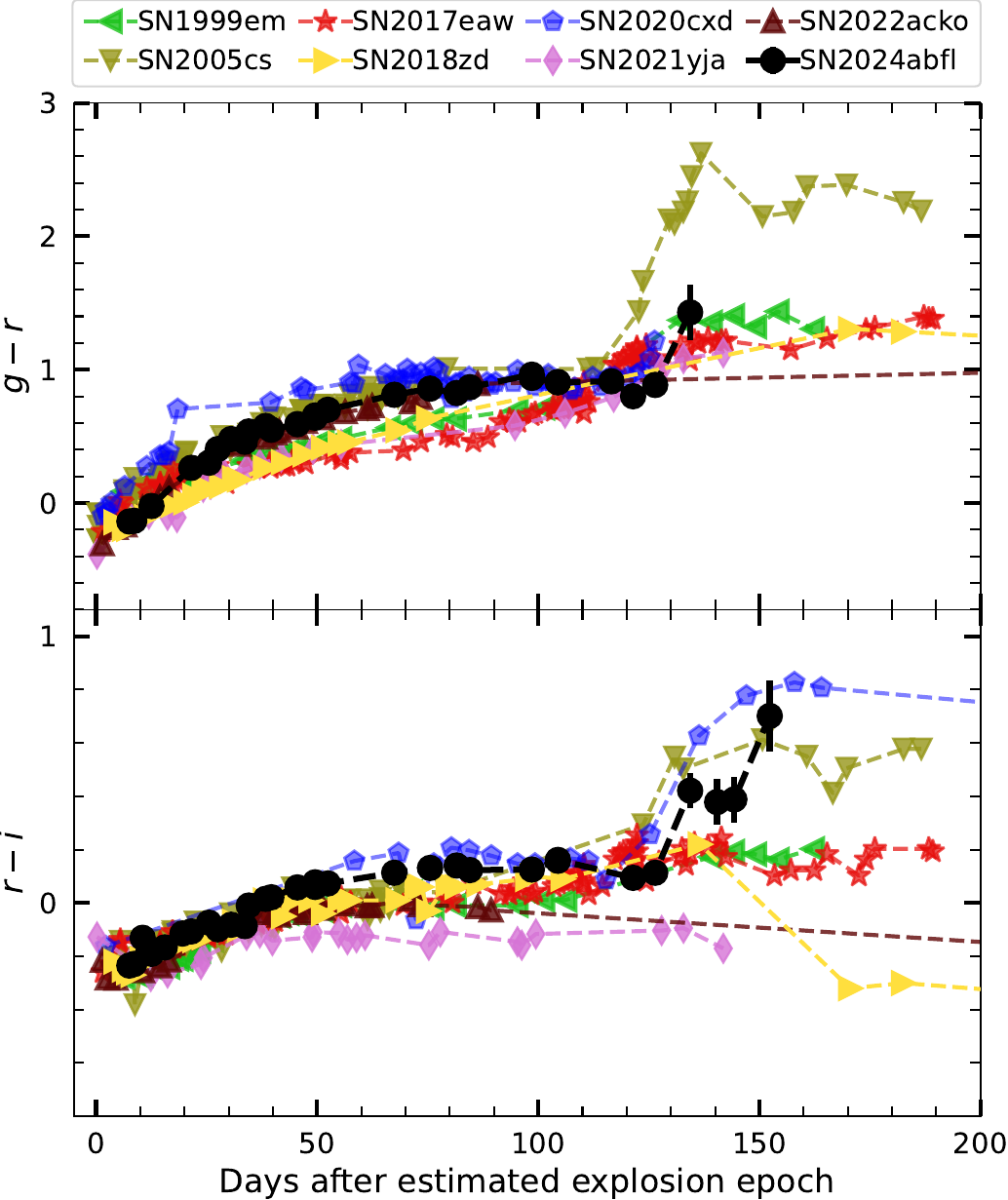}
   \caption{Colour evolution of SN\,2024abfl, along with that of other well-studied SNe II.
   All the colours have been corrected for both the Galactic and host-galaxy reddening.
   }
    \label{Fig:color}
\end{figure}  

Figure~\ref{Fig:color} presents the extinction-corrected $(g-r)$ and $(r-i)$ color of SN\,2024abfl, compared with those of a sample of typical SNe II listed in Table~\ref{tab:SNe_II info}.
Some SNe in the comparison sample lack Sloan-filter photometry, so we adopted the color conversion from \citet{2006A&A...460..339J} based on Johnson-filter data.
Overall, the color evolution of SN\,2024abfl is similar to that of other SNe,  particularly SN\,2005cs and SN\,2020cxd, although its colors are slightly redder than most SNe.
This is consistent with previous studies \citep{2014MNRAS.439.2873S,2009MNRAS.394.2266P} showing that LLSNe tend to have intrinsically redder colors compared to normal SNe II.

\begin{figure}
   \centering
   \includegraphics[width = 0.98\linewidth]{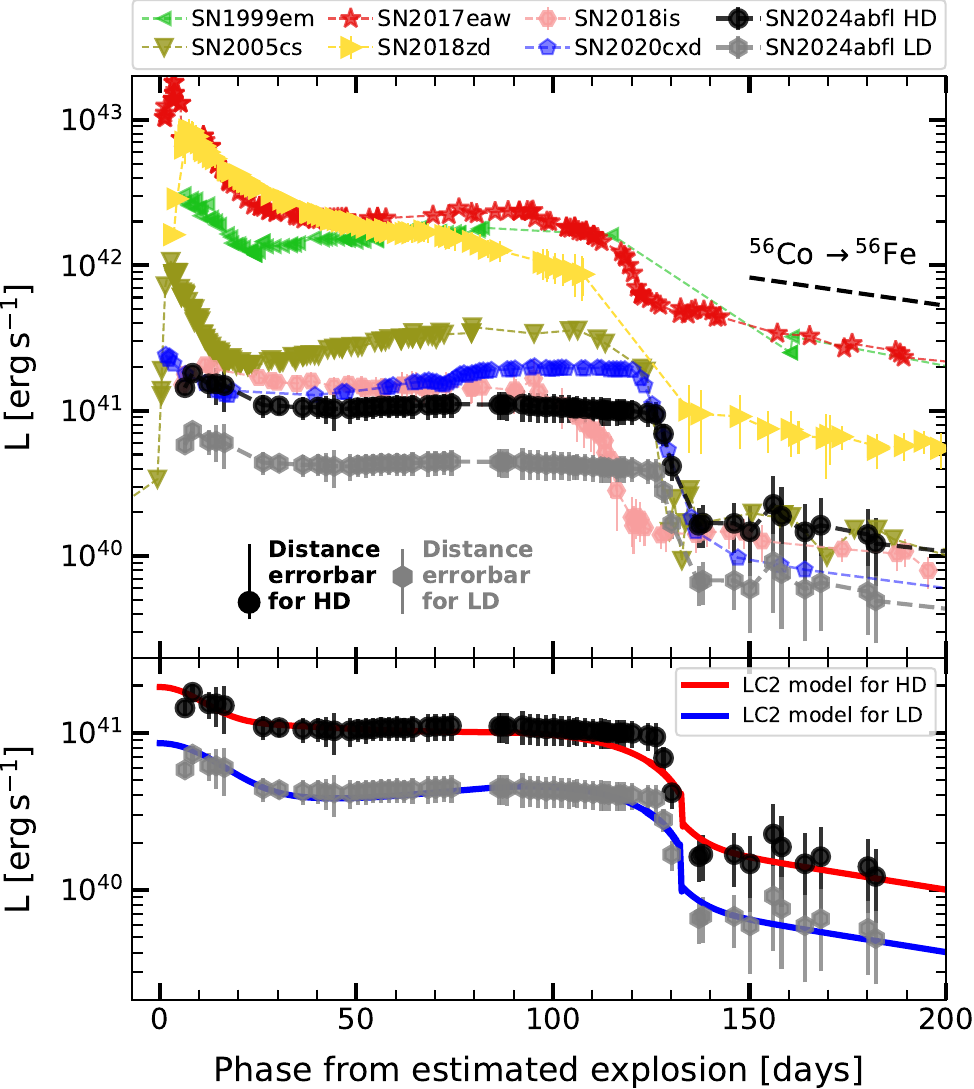}
   \caption{Bolometric luminosity evolution of SN\,2024abfl, along with that of other well-studied SNe II. 
   Due to the large distance uncertainty of SN\,2024abfl, we show only the luminosity error bar associated with the distance in the lower left of the top panel. 
   The bottom panel presents the two-component model fits to the SN\,2024abfl light curves, based on the model from \citet{2016A&A...589A..53N}.
   }
    \label{Fig:BL}
\end{figure}  

\begin{table}[!htbp]
    \centering
    \caption{Two-component model parameters of SN\,2024abfl based on different distance estimates. $R_0 $ is the initial radius of the ejecta, $M_{\mathrm{ej}}$ is the eject mass, $M_{\mathrm{Ni}}$ is nickel mass, $E_{\text {tot}}$ is the total energy including kinetic energy $E_{\text {kin}}$ and thermal energy $E_{\text{th}}$, $\kappa$ is the thomson scattering opacity.}
    \renewcommand{\arraystretch}{1.0}
    \setlength{\tabcolsep}{7pt}
    \begin{tabular}{ccccc}
        \hline \hline
        Parameter & \multicolumn{2}{c}{SN\,2024abfl HD} & \multicolumn{2}{c}{SN\,2024abfl LD}\\
        \hline
         & core & shell & core & shell \\ 
        \hline
        $R_0\ (10^{12} \mathrm{~cm})$ & 12 & 20 & 12 & 20 \\
        $M_{\mathrm{ej}}\ (\mathrm{M}_{\odot})$ & 6.0 & 0.3 & 4.0 & 0.3 \\
        $M_{\mathrm{Ni}}\ (\mathrm{M}_{\odot})$ & 0.0040 & -  & 0.0016 & - \\
        $E_{\text {tot }}\ (10^{51} \mathrm{erg})$ & 0.21 & 0.30 & 0.09 & 0.20  \\
        $E_{\mathrm{kin}} / E_{\mathrm{th}}$ &  2.4 & 50  & 3.8 & 50 \\
        $\kappa\ (\mathrm{~cm}^2 / \mathrm{g})$ & 0.23 & 0.40 & 0.30 & 0.40 \\
        \hline
    \end{tabular}
    \label{tab:LC2_parameters} 
\end{table}

\subsection{Bolometric light curve}
The bolometric light curve of SN\,2024abfl was constructed by using the photometric $gcVroiz$ bands.
The $u$ and $B$ bands lack mid-time observations, and their light curves decline faster than those of the other bands, making them unsuitable for interpolation.
The fitting procedure for the bolometric light curve employed the publicly accessible \texttt{Superbol}\footnote{\url{https://github.com/mnicholl/superbol}} programme \citep{2018RNAAS...2..230N}.
\texttt{Superbol} constructs the bolometric light curves by integrating extinction-corrected fluxes across observed bands.
Taking into account the reddening estimates and the two different distances reported in Section~\ref{sect:basic_inf}, we calculated the bolometric light curve of SN\,2024abfl and several SNe II listed in Table~\ref{tab:SNe_II info}, as shown in the top panel of Figure~\ref{Fig:BL}.

The bolometric light curve of SN\,2024abfl reveals a remarkably low luminosity compared to typical SNe IIP, such as SN\,1999em and SN\,2017eaw. 
Even within the sample of LLSNe, SN\,2024abfl remains at the lower end of the luminosity distribution. 
Its luminosity is comparable only to a few other well-studied low-luminosity SNe IIP, specifically SN\,2005cs, SN\,2018is, and SN\,2020cxd. Furthermore, when the lower distance estimate (derived using the SNe IIP distance method) is adopted, SN2024abfl’s bolometric luminosity is almost the lowest observed across the entire population of SNe IIP.
It is worth noting that the large distance uncertainty for SN\,2024abfl directly translates into significant uncertainty in its luminosity. 
Therefore, the comparison of SN\,2024abfl's luminosity to other SNe must be made with caution.

To obtain the explosion parameters of SN\,2024abfl, we applied the two-component light curve model of \citet{2016A&A...589A..53N} to fit the bolometric light curve of SN\,2024abfl based on different distances.
The model is based on a two-component configuration consisting of a dense inner He-rich region and an extended low-mass H-rich envelope for SNe IIP.
Here, the fitting results are shown in the bottom panel of Figure~\ref{Fig:BL}, and the corresponding parameters are listed in Table~\ref{tab:LC2_parameters}.
Most parameters of SN\,2024abfl are similar to the typical LLSNe, SN\,2005cs in \citet{2016A&A...589A..53N}.
However, the model fails to reproduce the sharp luminosity drop from the plateau to the radioactive tail, which is significantly steeper in the observations than predicted by the fit.
A similarly steep decline is also evident in the $Vroi$-band light curves.
The steep decline is more likely a direct consequence of the very low $^{56}$Ni mass synthesized in the explosion, which limits the radioactive heating supporting the recombination front \citep[e.g.,][]{2025MNRAS.538..223P}.
Additionally, stronger mixing of $^{56}$Ni into the outer ejecta may also produce a sharper drop from the plateau to the radioactive tail \citep[e.g.][]{2013MNRAS.434.3445P,2016ApJ...823..127N,2019MNRAS.483.1211K,2021ApJ...914....4U,2025arXiv251025018C}.
In the two-component model adopted here, the $^{56}$Ni energy production is assumed to be completely confined at the center \citep{2014A&A...571A..77N}.

The low luminosity observed in the late-time decay tail of SN\,2024abfl is directly responsible for the low inferred $^{56}$Ni mass. 
Based on the two distance estimates, $M_{\text{Ni}}$ is determined to be $0.004\,\rm M_\odot$ (for the high distance) and a remarkably low $0.0016\,\rm M_\odot$ (for the low distance).
SN\,2024abfl exhibits a similar and comparable tail luminosity to other well-studied LL SNe IIP, including SN\,2005cs, SN\,2018is, and SN\,2020cxd. 
For comparison, the nickel masses derived for these SNe are consistently low \citep[$\sim 0.003\,\rm M_\odot$; ][]{2009MNRAS.394.2266P,2021A&A...655A..90Y,2025A&A...694A.260D}.
\citet{2025arXiv250620068D} argued for a median nickel mass of $0.013^{+0.003}_{-0.010}\,\rm M_\odot$ and a range of $0.001-0.025\,\rm M_\odot$ for LLSNe IIP within the ZTF samples, based on fitting with radiation hydrodynamical models.
However, the lower distance estimate for SN\,2024abfl yields a nickel mass of $0.0016\,\rm M_\odot$, which places SN\,2024abfl at the extreme low end of the nickel mass distribution for SNe IIP.

We also utilized the formula provided by \citet{2003ApJ...582..905H} to calculate the radioactive nickel mass ($M_{\text{Ni}}$):
\begin{equation}
M_{\mathrm{Ni}}=\left(7.866 \times 10^{-44}\right) L_{\mathrm{t}} \times \exp \left[\frac{t /(1+z)-6.1}{111.26}\right] \mathrm{M}_{\odot}
\end{equation}
where $t$ is the phase after explosion, $L_{\rm t}$ is the tail-phase luminosity in units of $\rm erg\, s^{-1}$, and $z$ is the redshift of the SN.
Based on the tail luminosity values corresponding to the high distance and low distance estimates, the derived nickel masses $M_{\text{Ni}}$ are $ 0.005 \pm 0.002 \,\rm M_\odot$ and $0.002 \pm 0.001 \,\rm M_\odot$, respectively.
This $\rm ^{56}Ni$ mass is similar to that derived by the two-component model of \citet{2016A&A...589A..53N}, as both methods rely on a shared physical basis \citep{1982ApJ...253..785A,1989ApJ...340..396A}.

\begin{figure*}
   \centering
   \includegraphics[width = 0.8\textwidth]{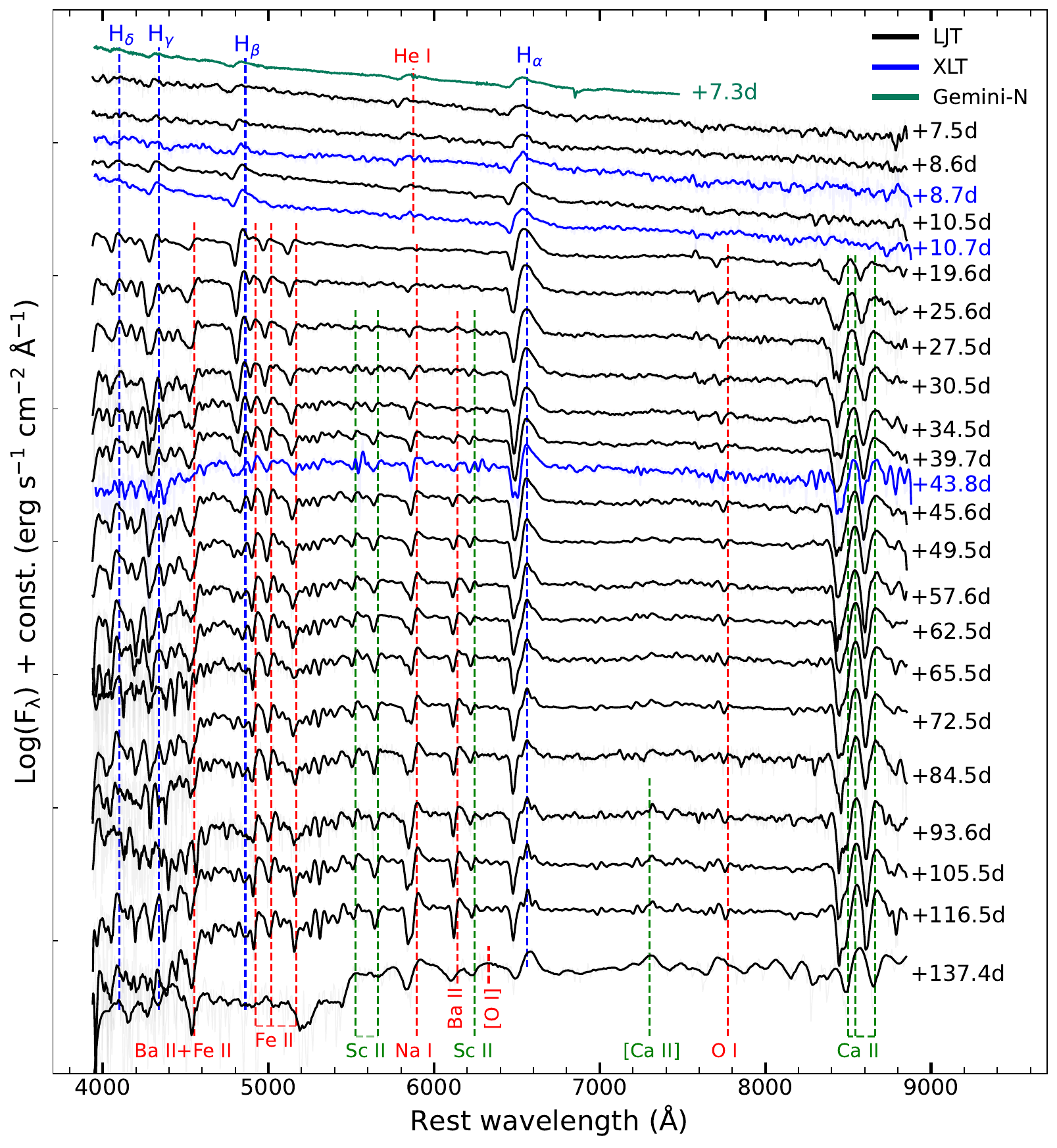}
   \caption{Spectral sequences of SN\,2024abfl.
    The positions of the principal transitions from H, He, and other spectral elements are highlighted by the dashed vertical lines. 
    The phase of spectra based on the estimated explosion epoch (MJD~=~60623.16) is given on the right-hand side.
    All spectra have been corrected for redshift and extinction.
    Most spectra, with lower S/N, have been smoothed using a Savitzky-Golay filter.}
    \label{Fig:spec}
\end{figure*}

\section{Spectroscopy}\label{sect:spectroscopy}

\begin{figure*}
   \centering
   \includegraphics[width = 0.9 \linewidth]{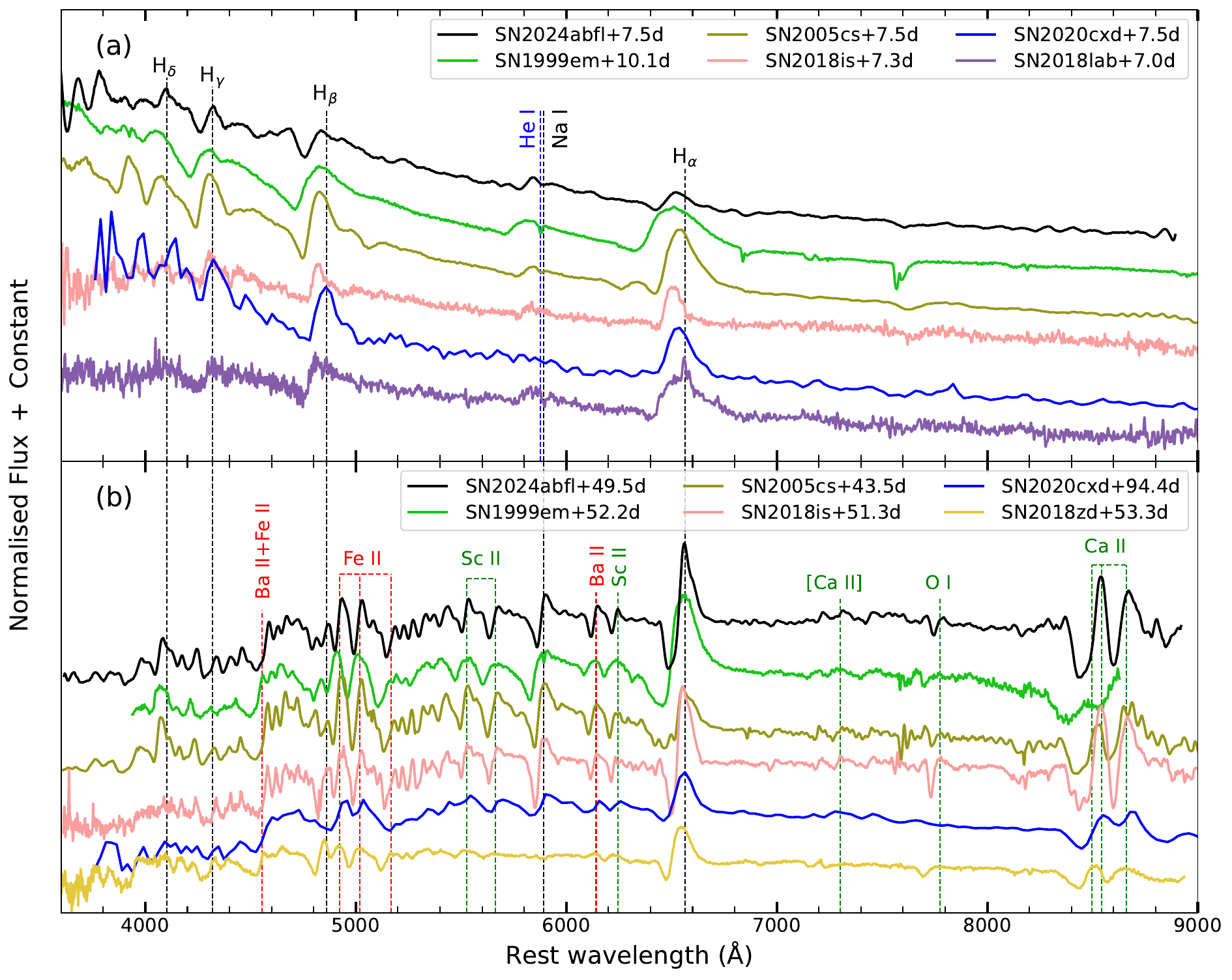}
   \caption{Early- and intermediate-phase spectroscopic comparisons of SNe IIP.
    The phases relative to the explosion date are indicated in the legend, and all spectra are shown in the rest frame.
    Key spectral features are marked.
    }
    \label{Fig:spec_com}
\end{figure*}

\begin{figure*}
   \centering
   \includegraphics[width = 0.9 \linewidth]{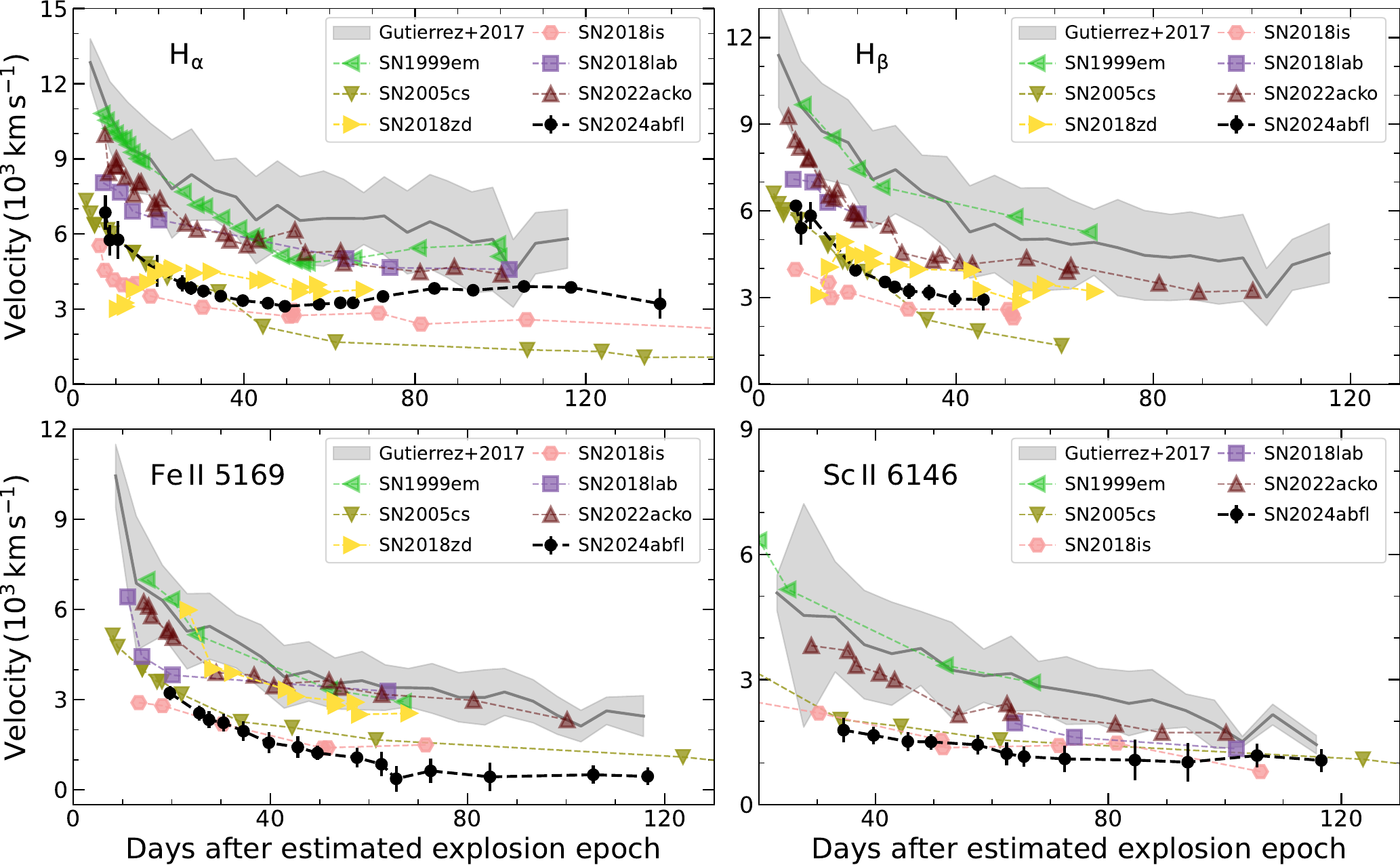}
   \caption{Line velocity evolution of \ha, \hb, \FeII~$\lambda\,5169$ and \ScII~$\lambda\,6246$ for SN\,2024abfl and comparisons with some typical samples, which are derived from the troughs of their P-Cygni absorption profiles. 
   The average velocity evolution of the SN II sample is shown as the gray region, taken from \citet{2017ApJ...850...89G}.
    Some of the velocities for the comparison SNe are adopted from \citet{2025A&A...694A.260D}.   
    }
    \label{Fig:line_velocity}
\end{figure*}

\subsection{Spectral sequence}

Figure~\ref{Fig:spec} presents the spectral evolution of SN\,2024abfl leading up to the onset of the nebular phase.
The first spectrum was obtained at +7.3\,d post explosion with Gemini-N, followed shortly by a second spectrum at +7.5\,d with LJT.
Up to +10.5\,d, the early-time spectra of SN\,2024abfl are characterized by a blue continuum with weak and relatively broad P-Cygni profiles of the hydrogen Balmer lines.
The first spectrum exhibits distinguishable \Naid absorption doublets originating from both the host galaxy and the Milky Way (see Figure~\ref{Fig:NaID}).
Additionally, a distinct feature around 5800\,\AA\ characterized by an emission peak near the \Naid absorption persists in the first five spectra. 
This feature is identified as the \Hei profile from the photospheric region, as illustrated in the zoomed-in view in Figure~\ref{Fig:spec_com}.

As the photosphere undergoes adiabatic expansion and cooling during the plateau phase, the continuum spectra temperature decreases.
This recombination phase is characterized by the emergence and strengthening of numerous narrow metal lines, including \FeII, \ScII, \Baii, \Oi, and \Caii. 
With time, these absorption features become narrower, deeper, and shift to redder wavelengths due to the recession of the photosphere into slower-moving ejecta. 
Notably, a distinct emission feature emerges within the broad \hb absorption trough in the spectra after +49.5\,d. 
This structure is likely attributed to the appearance of narrow metal lines, such as \Tiii\ and \FeII, which form around the broad \hb feature. 
The presence of these metal lines creates a complex absorption profile, visually mimicking an emission peak.

The final spectrum was obtained at +137.4\,d using the LJT.
By this epoch, SN\,2024abfl had already exited the recombination plateau phase, resulting in a significant drop in luminosity.
Consequently, this spectrum exhibits a relatively low S/N due to the faintness. 
Nevertheless, the forbidden emission lines of [\Oi] and [\Caii] are still identifiable in this spectrum.

\subsection{Comparison of SN IIP spectra}

Figure~\ref{Fig:spec_com} compares the early- and intermediate-time spectra of SN\,2024abfl with those of other SNe IIP, including SN\,1999em, SN\,2005cs, SN\,2018is, SN\,2020cxd, SN\,2018lab, and SN\,2018zd.
In panel~(a), we illustrate the early spectrum of SN\,2024abfl on +7.5\,d together with those of other SNe IIP at comparable epochs.
These early spectra exhibit a blue continuum, consistent with a high blackbody temperature shortly after explosion. 
The continuum and P-Cygni profiles of SN\,2024abfl closely resemble those of SN\,2008is.
Compared with other objects in the sample, both SNe likely exhibit somewhat lower temperatures and line velocities.
In addition, they also display prominent features of \ha, \hb, \hg, and \hd. 
The P-Cygni profile around 5800\,\AA, including a narrow absorption component, is attributed to \Hei and \Naid, except in the case of SN\,2020cxd where the limited S/N obscures this feature. 

The spectrum of SN\,2024abfl at $\sim$50\,d post-explosion is also compared to the other SNe IIP at similar epochs, as shown in panel~(b) of Figure~\ref{Fig:spec_com}.
Although the \ha P-Cygni features are present in all spectra at this phase, their strength and width vary significantly among different SNe IIP.
For SN\,2024abfl, the \ha feature is relatively narrow, comparable to that of SN\,2018is.
In addition to the hydrogen lines, a variety of complex metal features appear in the spectra at this epoch, including \Feii, \Baii, \Scii, \Caii, and \Oi, particularly toward the blue end of the spectrum.
Overall, the plateau-spectrum morphology of SN\,2024abfl closely resembles that of the prototypical LL SN\,2005cs and SN\,2018is.
The presence of narrow, well-defined \Baii and \Scii lines is a common feature in LL SNe IIP. 
In typical SNe IIP, higher ejecta velocities generally broaden and blend these lines.

\subsection{Spectra line velocity evolution}

The velocity evolution of several spectral lines, including \ha, \hb, \FeII~$\lambda\,5169$ and \ScII~$\lambda\,6246$, is presented in Figure~\ref{Fig:line_velocity}.
The line velocities are determined from the absorption minima of P-Cygni profiles, using MCMC-based fitting with a two-component Gaussian profile.
For comparison, we also include the SNe II sample from \citet{2017ApJ...850...89G} as well as several other well-observed SNe II.

The expansion velocities of SN\,2024abfl exhibit a clear evolution characteristic of LL SNe IIP.
In the early phase ($< 40\,$d), the \ha velocity of SN\,2024abfl is similar to that of the prototypical LL SN\,2005cs, though slightly higher than SN\,2018is. 
After 40\,d post-explosion, the \ha velocity settles at approximately $3500\,\kms$ with minimal evolution thereafter.
The early-time \hb velocity of SN\,2024abfl is initially comparable to SN\,2005cs. 
We note that it is complicated to estimate late-time \hb velocity due to the intricate morphology of the P-Cygni absorption feature.
Furthermore, the \FeII and \ScII velocities of SN\,2024abfl are significantly lower than those of the comparison sample, including SN\,2005cs and SN\,2018is.
\citet{2024MNRAS.528.3092L} presented a spectral data release for 104 SNe II, showing that the \FeII velocities at $t\sim 50$ d span a range of 2000–5500$\,\kms$, with a mean value of $3872 \pm 949\,\kms$.
The \FeII velocity of SN\,2024abfl at $\sim$50\,d is approximately $1200\,\kms$, which falls well below this entire distribution.
Collectively, these exceptionally low line velocities reinforce the classification of SN\,2024abfl as an LL SN IIP, consistent with a low-energy explosion from a low-mass progenitor.

\section{Discussion} \label{sect:discussion}

\subsection{Progenitor constrains} \label{sect:progenitor}

\begin{figure}
   \centering
   \includegraphics[width = \linewidth]{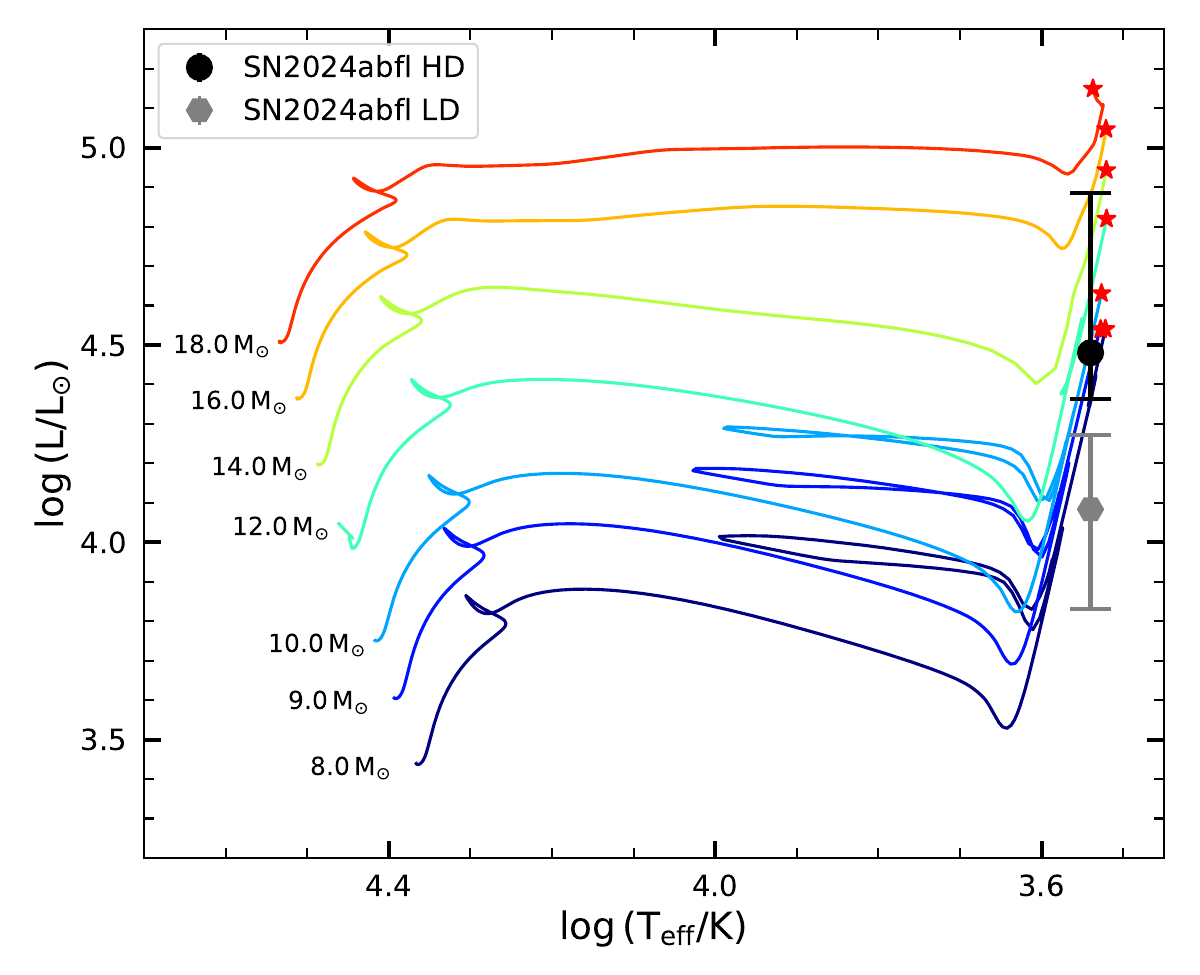}
      \caption{MIST evolutionary tracks for massive stars with initial masses ranging from 8 to 18$\,\rm M_\odot$, shown alongside the progenitor position of SN\,2024abfl. 
      The red star symbols indicate the points of core carbon depletion.
      This figure is adapted from \citet{2025ApJ...982L..55L}.} 
     \label{fig:progenitor}
\end{figure}

\citet{2025ApJ...982L..55L} recently identified the progenitor of SN\,2024abfl in archival HST data, suggesting an RSG star with an initial mass of $9$-$12\,\rm M_\odot$ (assuming a distance of $15.6^{+6.1}_{-3.0}\,$Mpc).
Given the complex interactions within the host galaxy, we examined nearly all available distance measurements and identified two distinct distance distributions, as detailed in Section~\ref{sect:basic_inf}. 
Utilizing these two distance estimates, we placed the progenitor on the H–R diagram, as shown in Figure~\ref{fig:progenitor}. 
For comparison, we also plotted MIST evolutionary tracks\footnote{\url{https://waps.cfa.harvard.edu/MIST/interp_tracks.html}} for massive stars with an initial mass range of $8-18\,\rm M_\odot$ \citep{2016ApJS..222....8D,2016ApJ...823..102C,2011ApJS..192....3P,2013ApJS..208....4P,2018ApJS..234...34P}. 
It is worth noting that here our adopted pre-explosion extinction ($A_{\rm F814W} = 1$) and progenitor HST F814W magnitude are adopted from  \citet{2025ApJ...982L..55L}.

Adopting the higher distance estimate for SN\,2024abfl, the inferred progenitor luminosity implies an initial mass of $8-12\,\rm M_\odot$, consistent with previous results derived using a similar distance \citep{2025ApJ...982L..55L}.
However, under the lower distance assumption, the progenitor luminosity falls significantly below the terminal points of standard evolutionary tracks (i.e., $<8\,\rm M_\odot$).
We propose these possible scenarios to explain this discrepancy. 
First, the progenitor may have suffered from significant self-obscuration by circumstellar dust during the pre-explosion RSG phase, leading to an underestimation of the intrinsic luminosity.
If the progenitor luminosity is required to be consistent with that expected for low-mass RSGs under the lower distance estimate, a larger extinction, with $A_{\rm F814W} \gtrsim 1.7$, would be required.
Second, the lower distance derived from SN\,IIP-based methods might be unreliable.
As noted by \citet{2016A&A...588A...1P,2021arXiv210912943C}, peculiar SNe such as LSQ13fn and SN\,2018zd do not necessarily follow the standardised candle method (SCM) commonly used for distance determination. 
In fact, \citet{2020MNRAS.494.5882R} identify a subset of SNe II—including SN\,2008bm, SN\,2009aj, SN\,2009au, SN\,1983K, and LSQ13fn—as luminous SNe II with low expansion velocities (LLEV), which can be reproduced assuming ejecta–CSM interaction lasting between 4 and 11 weeks post-explosion. 
\citet{2021arXiv210912943C} also note that some features of SN\,2018zd resemble LLEV SNe.
It is worth noting that \citet{2026arXiv260102638G} detects a broad ledge feature around 4600\AA within three days of explosion for SN\,2024abfl, corresponding to a blend of high-ionization, shock-accelerated CSM lines, even though SN\,2024abfl exhibits a lower absolute magnitude. 
Therefore, SN IIP-based distance methods may not be suitable for LLEV SNe or SNe with prominent ejecta–CSM interaction.
Finally, the progenitor of SN\,2024abfl could have undergone complex binary evolution stage, rendering standard single-star evolutionary tracks inapplicable.

\subsection{Comparisons with SN samples}
\begin{figure*} 
   \centering
   \includegraphics[width = 0.9\textwidth]{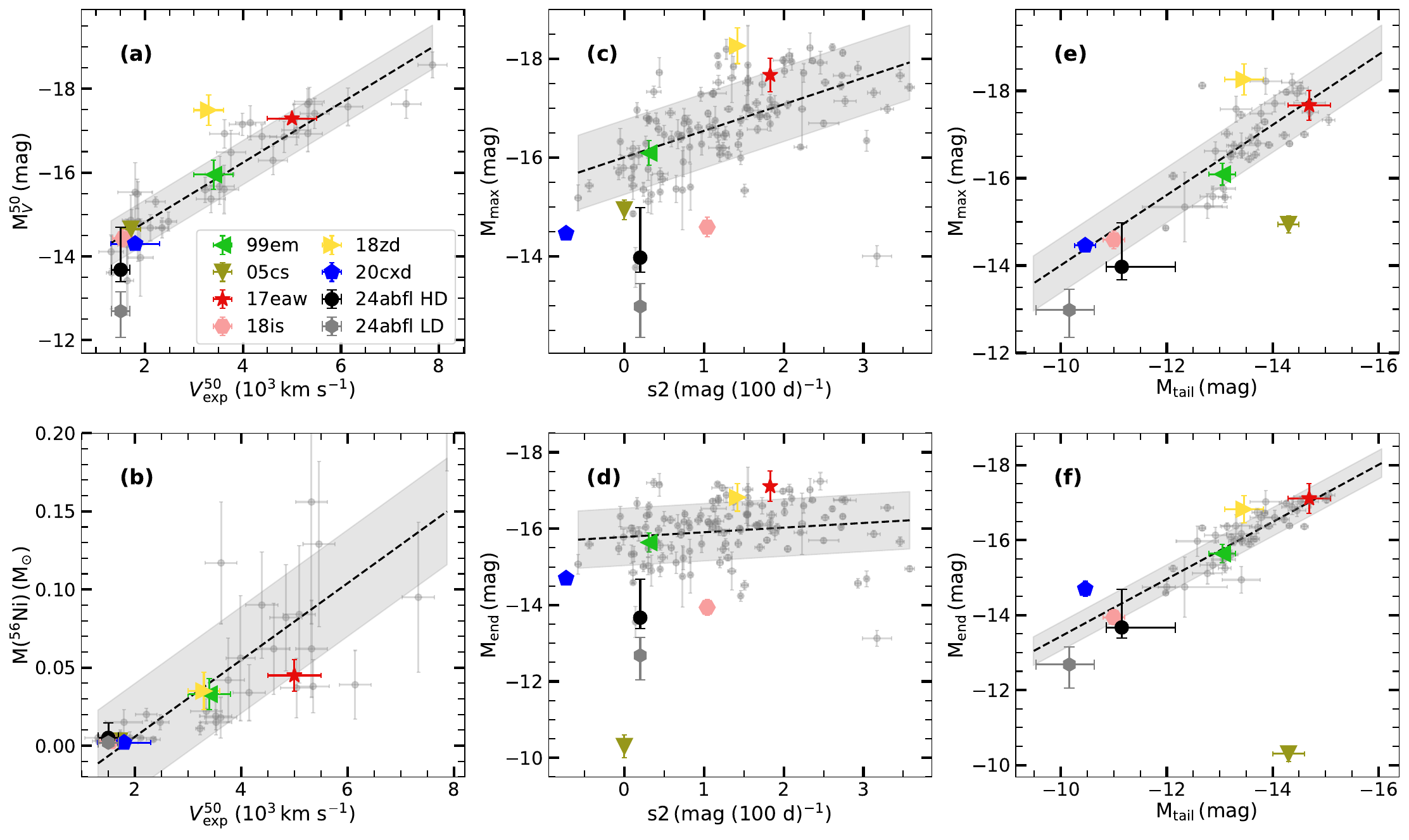}
   \caption{The position of SN\,2024abfl in the SN II samples considering various photometric and spectroscopic indicators, including the velocity of \ScII measured at $t\approx 50\,$d after the explosion ($v_{\rm exp}^{50}$), the V-band absolute magnitude measured at at $t\approx 50\,$d after the explosion ($\mathrm{M}_{V}^{50}$), the mass of $^{56}$Ni ($\rm M(^{56}Ni)$), and four shape parameters of the $V$-band light curve (as defined by \citet{2014ApJ...786...67A} – the absolute brightness at three selected phases ($\rm M_{max}$, $\rm M_{end}$, and $\rm M_{tail}$) and the decline rate of the plateau (s2).
    For SNe lacking $V$-band observations, the corresponding $r$-band data were adopted as a replacement.
   The comparison sample used for the correlation analysis is shown as gray open circles. 
   This data set comprises SNe II observations collected from the literature \citep{2003ApJ...582..905H, 2014ApJ...786...67A,2014MNRAS.439.2873S, 2014ApJ...797....5Z}. 
   Black dashed lines represent the linear fits derived from this comparison sample, while the gray shaded field illustrates the $1\sigma$ uncertainties associated with these linear fits.
   For comparison, several typical SNe IIP (SN\,1999em, SN\,2005cs, SN\,2017eaw, SN\,2018is, SN\,2018zd, and SN\,2020cxd) are also included in the figure.}
    \label{Fig:comparison}
\end{figure*}  

Figure~\ref{Fig:comparison} illustrates the location of SN\,2024abfl in the parameter space of SNe II based on various spectroscopic and photometric indicators.
The two distance estimates and associated uncertainties for SN\,2024abfl are also adopted here.
Compared with the linear fitting in each panel, the peak and end magnitudes of SN\,2024abfl are lower than those of most comparison SNe, and SN\,2024abfl lies below the majority of fitted relations, even when the higher distance is adopted.
However, SN\,2024abfl stands out as an outlier in the plateau decline rate–magnitude plane, as shown in Panels (c) and (d). 
For SNe with comparable plateau decline rates, the peak and end absolute magnitudes of SN\,2024abfl are significantly fainter than those of the majority of events, similar to the behavior seen in SN\,2018is and SN\,2005cs.
Overall, SN\,2024abfl resembles typical LL SNe IIP, such as SN\,2005cs, SN\,2018is, and SN\,2020cxd, particularly in its low envelope velocity, low plateau magnitude, and low $^{56}$Ni mass, all of which are characteristic of this subclass.
In these panels, the position of SN\,2024abfl derived from the higher distance estimate appears to align more closely with the parameters of LL SNe IIP than that derived from the lower distance, suggesting that the higher estimate is likely closer to the true distance of the host galaxy.
Additionally, SN\,2018zd exhibits a higher luminosity and occupies a position offset from the bulk of the most samples in Panels (a), (c), and (e). 
The parameter-space locations of SN\,2018zd are also well separated from those of SN\,2005cs and SN\,2024abfl.

\subsection{Formation scenario and distance }

ECSNe from SAGB progenitors are expected to appear as relatively faint SN II (IIP or IIL) or stripped-envelope SNe \citep{2009ApJ...705L.138P,2014A&A...569A..57M,2015ApJ...810...34W}.
However, it remains challenging to observationally distinguish between LL SNe IIP arising from low-mass CCSNe and those from ECSNe.
Recently, \citet{2024ApJ...970..163S} proposed a light-curve diagnostic to discriminate between these two classes, noting that the SNe IIP from ECSNe typically exhibit bluer plateau phases.
They suggested the following sufficient condition for an ECSN identification:
\begin{equation}
(g-r)_{t_{\mathrm{PT}} / 2}<0.008 \times t_{\mathrm{PT}}-0.4
\end{equation}
where $(g-r)_{t_{\mathrm{PT}} / 2}$ denotes the extinction-corrected color at the mid-point of the plateau phase ($t = t_{\mathrm{PT}} / 2$).
Applying this method to a sample of SNe IIP, \citet{2024ApJ...970..163S} classified SN\,2018zd as an ECSN.
For SN\,2024abfl, we measure $(g-r)_{t_{\mathrm{PT}} / 2} \approx 0.8$. Given its plateau duration, the threshold value is about $0.6$.
Since SN\,2024abfl does not satisfy this ECSN criterion, this diagnostic favors a low-mass CCSN classification.

On the other hand, \citet{2013ApJ...771L..12T} suggested that the $V$-band light curve of ECSNe is characterized by a significant luminosity drop of $\sim4\,$mag from the plateau to the radioactive tail. 
The drop observed in SN\,2018zd is approximately $3.8\,$mag, which is consistent with the ECSN prediction. 
However, a similarly large $V$-band drop of $\sim4$ mag is also observed in the LL SN\,2005cs \citep{2009MNRAS.394.2266P}.
In contrast, LL SN\,2024abfl exhibits a much smaller drop of $\sim 2.4\,$mag, more consistent with typical SNe IIP.
These comparisons indicate that the magnitude drop from the plateau to the tail is likely degenerate and cannot serve as a robust discriminator between ECSNe and other SNe II. 
Indeed, some LL SNe may still be compatible with an ECSN scenario, and their formation channels remain under debate.

Furthermore, the plateau duration of ECSNe is predicted to be $60-100\,$days \citep{2013ApJ...771L..12T}, which is relatively shorter than that of LL SNe IIP. 
In contrast, SN\,2024abfl exhibits a durable plateau that is longer than that of most SNe IIP, as shown in Figure~\ref{fig:Abs_r}.
Additionally, the spectroscopic evolution and parameter features of SN\,2024abfl closely resemble those of the LL SN\,2005cs and SN\,2018is, as shown in Figure~\ref{Fig:spec_com}.
These two LL SNe IIP have commonly been suggested to originate from the low-mass CCSN scenario \citep[e.g.,][]{2009MNRAS.394.2266P,2025A&A...694A.260D}.

In summary, we suggest that the collective observational properties of SN\,2024abfl favor a very low-mass CCSN origin.
However, considering the limitations of current theoretical models and the existence of multiple potential formation channels for ECSNe, we cannot definitively rule out the ECSN scenario.

The distance to NGC\,2146 has long been subject to significant uncertainty, largely owing to the galaxy’s complex tidal interactions, which has also hindered detailed interpretation of SN\,2018zd.
In this work, our examination of the available distance estimates reveals a bimodal distribution, and we therefore consider both distance solutions in the relevant analyses.
Based on SN sample statistics and the inferred progenitor luminosity, we argue that the larger distance is more plausible, as it places SN\,2024abfl more naturally within a low-mass CCSN scenario.
For SN\,2018zd, the true distance likewise appears to favor the larger-distance solution, consistent with the conclusions of \citet{2020MNRAS.498...84Z} and \citet{2021arXiv210912943C}. 
Under this assumption, the synthesized $^{56}$Ni mass would be higher, reaching $\rm \sim 0.02$–$0.03\,M_\odot$.
Moreover, \citet{2021arXiv210912943C} reported that the Ni/Fe ratio in SN\,2018zd is significantly lower than expected for ECSNe.
Taken together, these considerations suggest that the formation channel of SN\,2018zd remains uncertain and warrants further detailed investigation.

\section{Conclusion} \label{sect:conclusion}

We presented the discovery and follow-up observations of the LL SN IIP SN\,2024abfl, characterized by a long-lasting plateau phase and unusually low expansion velocities. 
The event is projected near the position of SN\,2018zd, previously proposed as a possible ECSN candidate \citep{2021NatAs...5..903H}.
Its host, the starburst galaxy NGC\,2146, exhibits disturbed morphology likely driven by tidal interactions with a low–surface-brightness companion \citep{2001A&A...365..360T}, which complicates distance determinations and leads to discrepant results across methods \citep{2021arXiv210912943C}.
Our examination of the available distance estimates reveals a bimodal distribution: 
the shorter distance is primarily derived from SN\,IIP EPM and SCM methods, whereas the larger distance comes from kinematic measurements, the Tully–Fisher relation, and globular cluster analyses. 
Significant uncertainties remain in all these methods owing to the disturbed structure and dynamics of the host galaxy.
A more reliable and precise distance determination (e.g., TRGB method) is therefore required to firmly establish the true distance to NGC\,2146.

Under both distance scales, SN\,2024abfl is intrinsically faint. Its absolute plateau magnitude and bolometric luminosity lie well below those of typical SNeIIP; even at the larger distance (including uncertainties), the event remains at the faint end of the LL IIP population. 
Modeling the bolometric luminosity evolution with a two-component framework indicates an unusually small synthesized $^{56}$Ni mass, lower than that of most LL SNe IIP.
Spectroscopically, SN\,2024abfl closely resembles LL events such as SN\,2005cs and\,SN 2018is.
Its \Feii and \Scii line velocities are among the lowest in our comparison sample, reinforcing the picture of a low-energy explosion.
Given its extremely low expansion velocities, intrinsically faint plateau, and unusually low $^{56}$Ni mass, SN\,2024abfl appears to be an exceptionally LL SN IIP.
Considering the plateau color and duration, the magnitude drop between plateau and tail, and the spectroscopic properties, the event is more consistent with a low-mass CC origin than with an ECSN scenario. 
A robust distance measurement will be crucial to refine the explosion parameters and further test this interpretation.

\section*{Acknowledgements}
We gratefully thank the anonymous referee for his/her insightful comments and suggestions that improved the paper.
This study is supported by the CAS Project for Young Scientists in Basic Research (YSBR-148), 
the National Natural Science Foundation of China (Nos 12225304, 12288102), the National Key R\&D Program of China (No. 2021YFA1600404), the Yunnan Revitalization Talent Support Program (Yunling Scholar Project), the Yunnan Science and Technology Program (Nos 202501AS070005, 202605AS350010), and the International Centre of Supernovae (ICESUN), Yunnan Key Laboratory of Supernova Research (No. 202505AV340004). 
J.Z. is supported by the B-type Strategic Priority Program of the Chinese Academy of Sciences (Grant No. XDB1160202), the National Natural Science Foundation of China (NSFC grants 12173082 and 12333008),  the Yunnan Fundamental Research Projects (YFRP; grants 202501AV070012 and 202401BC070007).
The work of X.-F.W. is supported by the National Natural Science Foundation of China (NSFC grants 12288102 and  12033003), the Ma Huateng Foundation, and the New Cornerstone Science Foundation through the XPLORER PRIZE.

We thank Raya Dastidar for the helpful assistance and for providing the comparison data.
We acknowledge the support of the staff of the LJT and XLT.
Funding for the LJT has been provided by the CAS and the People’s Government of Yunnan Province. 
The LJT is jointly operated and administrated by YNAO and Center for Astronomical Mega-Science, CAS.

This work made use of data obtained from the Swift satellite, including observations from the Ultraviolet/Optical Telescope (UVOT). These data were supplied by the UK Swift Science Data Centre at the University of Leicester.

Based on observations obtained with the Samuel Oschin Telescope 48-inch and the 60-inch Telescope at the Palomar
Observatory as part of the Zwicky Transient Facility project. ZTF is supported by the National Science Foundation under Grants No. AST-1440341 and AST-2034437 and a collaboration including current partners Caltech, IPAC, the Oskar Klein Center at Stockholm University, the University of Maryland, University of California, Berkeley, the University of Wisconsin at Milwaukee, University of Warwick, Ruhr University, Cornell University, Northwestern University and Drexel University. Operations are conducted by COO, IPAC, and UW.

This work has made use of data from the Asteroid Terrestrial-impact Last Alert System (ATLAS) project. The Asteroid Terrestrial-impact Last Alert System (ATLAS) project is primarily funded to search for near earth asteroids through NASA grants NN12AR55G, 80NSSC18K0284, and 80NSSC18K1575; byproducts of the NEO search include images and catalogs from the survey area. This work was partially funded by Kepler/K2 grant J1944/80NSSC19K0112 and HST GO-15889, and STFC grants ST/T000198/1 and ST/S006109/1. The ATLAS science products have been made possible through the contributions of the University of Hawaii Institute for Astronomy, the Queen’s University Belfast, the Space Telescope Science Institute, the South African Astronomical Observatory, and The Millennium Institute of Astrophysics (MAS), Chile.

\bibliography{24abfl}{}
\bibliographystyle{aasjournalv7}



\appendix
\small

\restartappendixnumbering

\section{Supplementary figures}  \label{sect:supfigure}

\begin{figure}
   \centering
   \includegraphics[width = 0.45\linewidth]{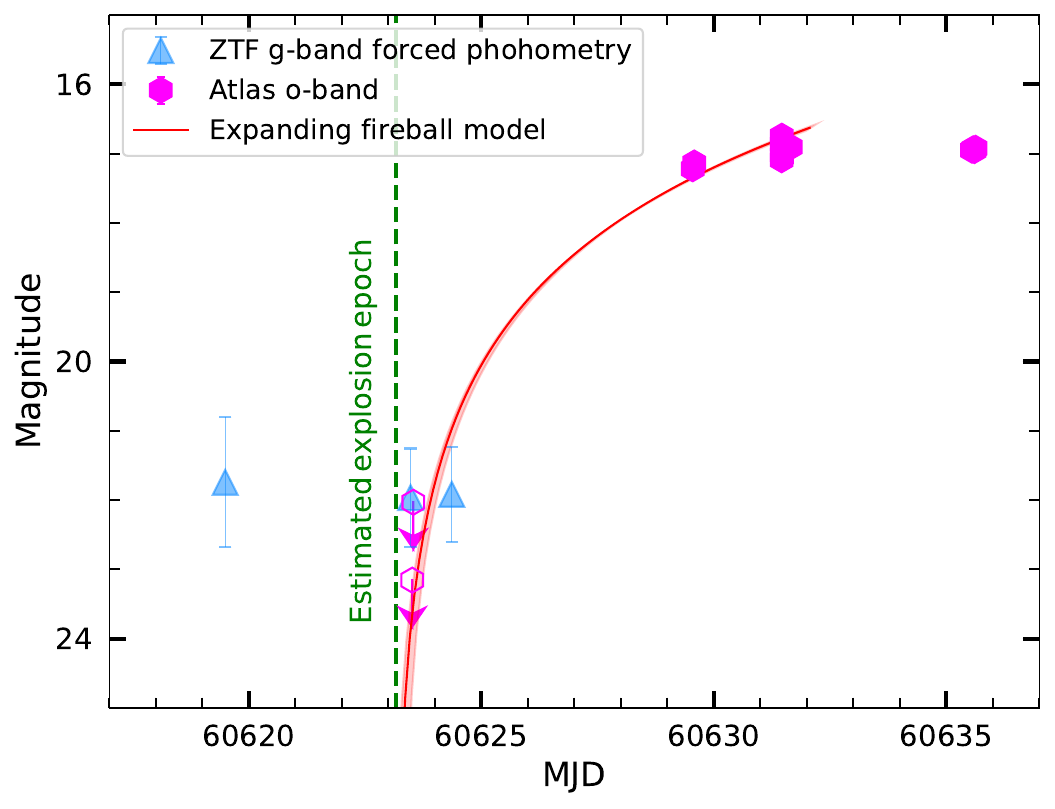}
      \caption{Expanding fireball model fit to the early detections of SN\,2024abfl in the ATLAS $o$ band and ZTF $g$ band. 
      The first ZTF detection is excluded from the fit. 
      Note that the ATLAS data are not stacked, while the ZTF data are from forced photometry.}
     \label{Fig:explosion_epoch}
\end{figure}

\begin{figure}
\includegraphics[width=0.48\textwidth]{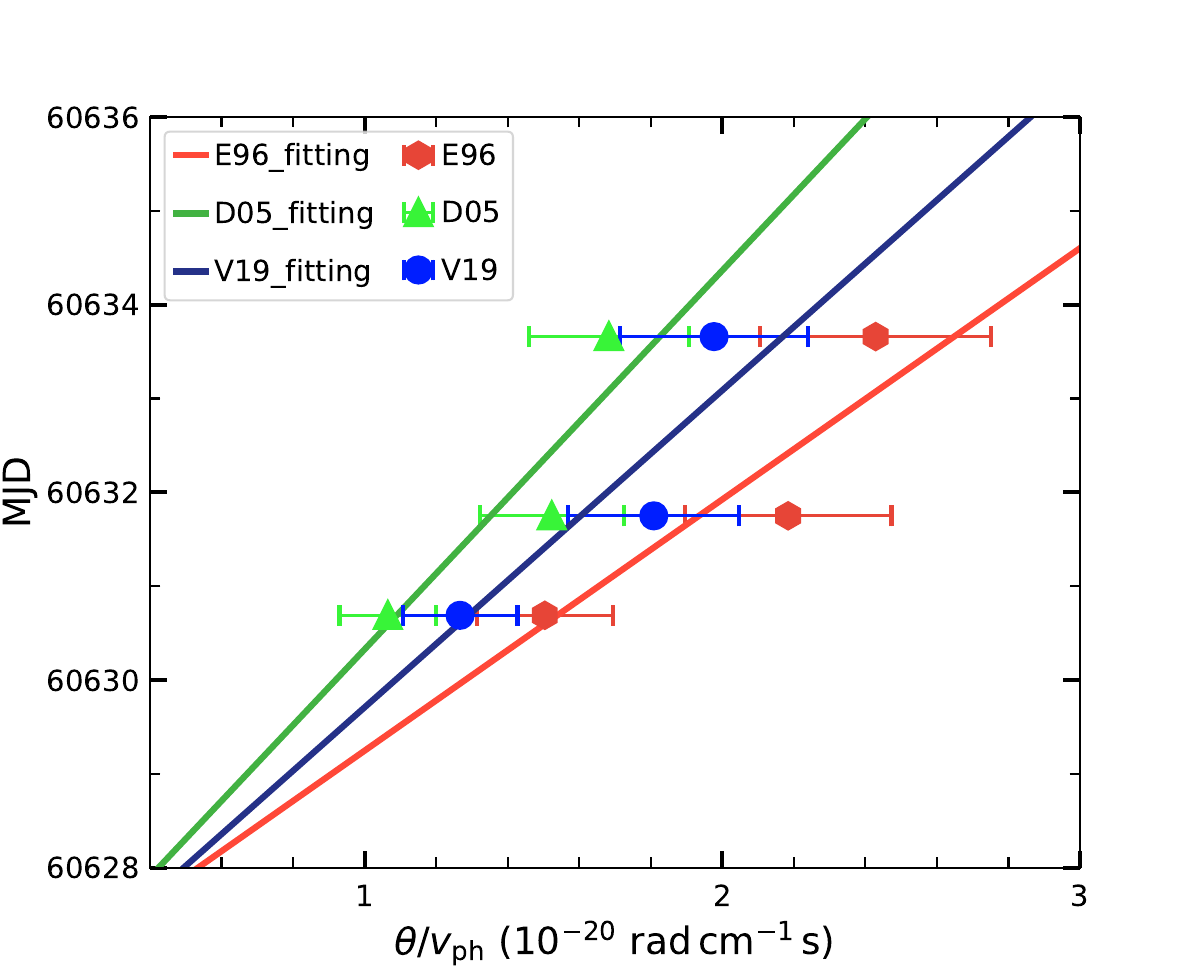}
\includegraphics[width=0.44\textwidth]{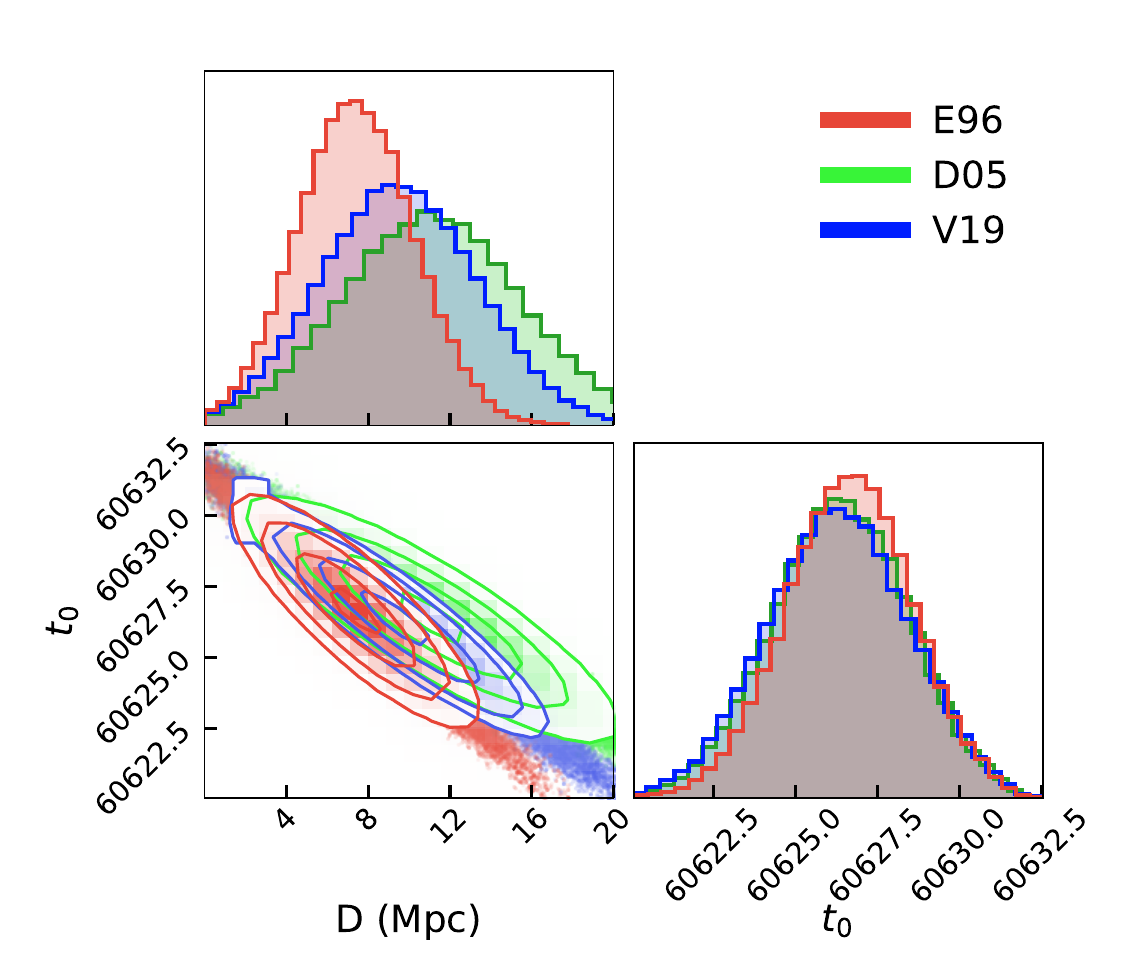}
    \caption{Linear fit to explosion epoch ($t_0$) and $\theta/v_{\rm ph}$ to determine the explosion epoch and distance.
    The left panel shows the linear fit for the three methods based on $BV$-band  photometric data.  
    The right panel displays the one- and two-dimensional projections of the posterior probability distributions for slope (D) and intercept $t_0$ across the three method sets in the corner plot.}
    \label{fig:EPM_fitting}
\end{figure}

\begin{figure}
   \centering
   \includegraphics[width = 0.5\linewidth]{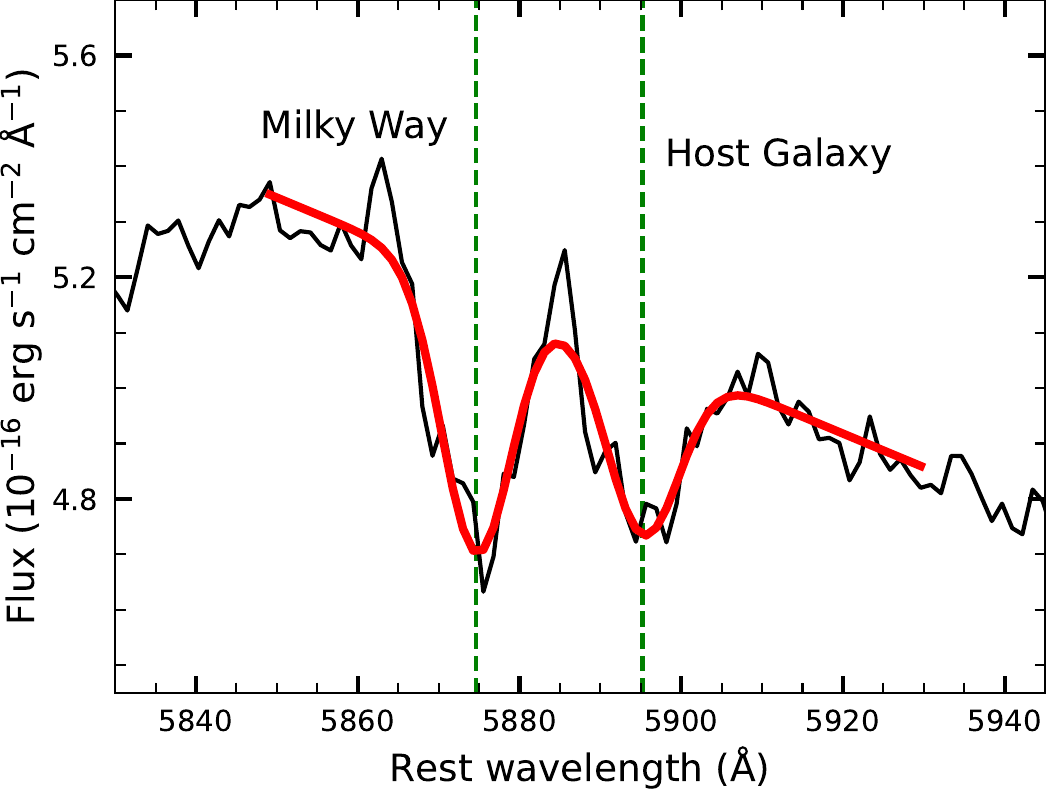}
      \caption{\Naid absorption features and corresponding fitting results for SN\,2024abfl from the Gemini-N/GMOS spectrum obtained on 2024-11-16.}
     \label{Fig:NaID}
\end{figure}

\begin{figure*}
\centering
\includegraphics[width=0.444\textwidth]{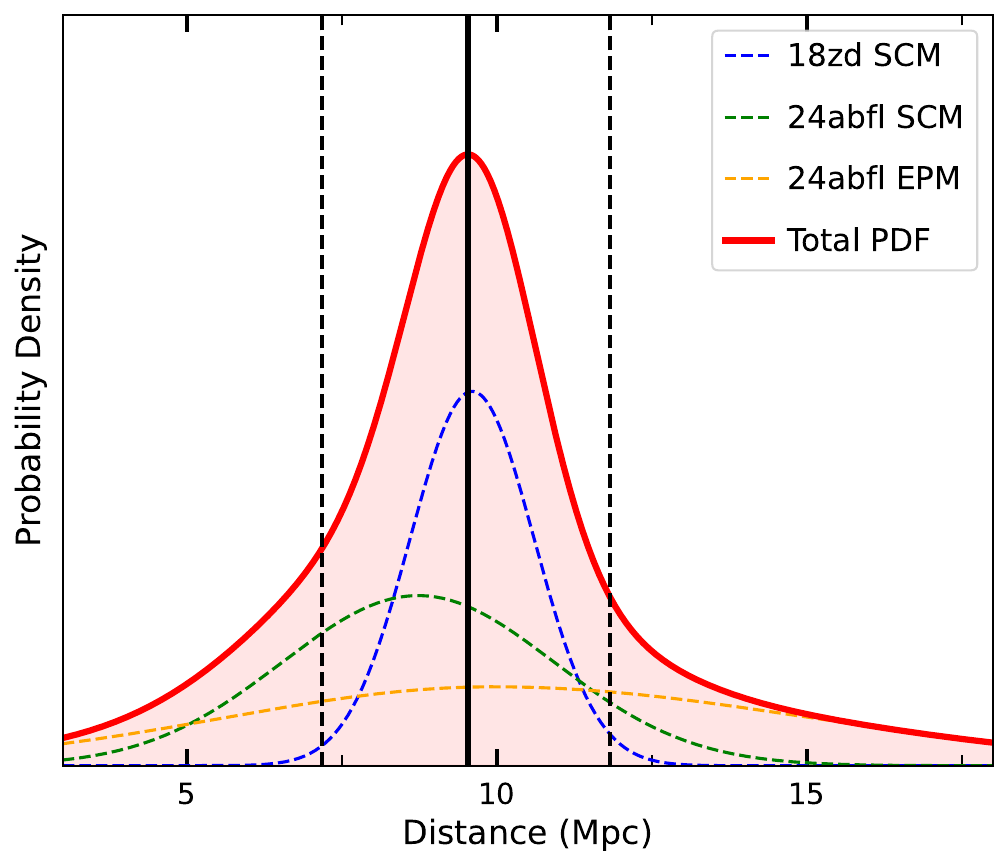}
\includegraphics[width=0.45\textwidth]{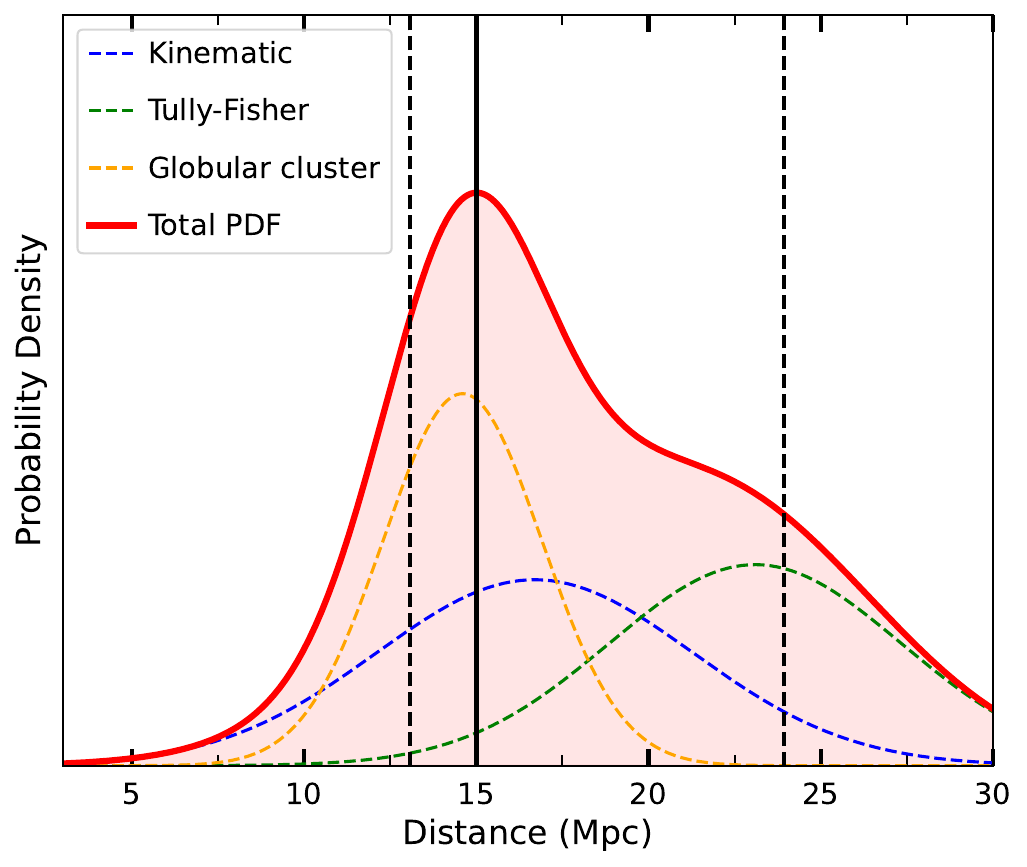}
\caption{PDFs for the different distance measurements are shown, together with the joint PDF for NGC\,2146 (solid red line). 
The solid black vertical line marks the maximum of the joint PDF, while the dashed vertical lines indicate the 16th and 84th percentiles.
In the left panel, we adopted the distances derived from the SCM of SN\,2018zd and SN\,2024abfl, as well as the EPM distance of SN\,2024abfl.
In the right panel, we performed the distances based on the kinematic method, the CO-line Tully–Fisher relation \citep{1997A&A...326..915T}, and the globular cluster distance \citep{2021arXiv210912943C}.
}
\label{fig:PDFs}
\end{figure*}

In this appendix, we present the fitting results of the expanding fireball model to obtain a more accurate explosion epoch, as shown in Figure~\ref{Fig:explosion_epoch}. 
We also apply the expanding photosphere method (EPM) to SN\,2024abfl for distance determination, and the corresponding fitting results are shown in Figure~\ref{fig:EPM_fitting} and Table~\ref{table:EPM quantities}. 
Additionally, the earliest spectrum exhibits a distinct \Naid absorption feature arising from both the Milky Way and the host galaxy, as shown in Figure~\ref{Fig:NaID}. 
Based on the equivalent widths (EWs) of this host-galaxy feature, we estimate a host-galaxy extinction of $E(B-V)_{\rm host}=0.10\,$mag. 
Finally, we compile different distance measurements and construct two probability density functions (PDFs) to constrain the final distance to the host galaxy, as shown in Figure~\ref{fig:PDFs}.

\section{Supplementary tables} \label{sect:suptable}

Here, the corresponding multiband photometric data and spectroscopic observation logs are presented in Tables~\ref{table:photdata} and \ref{table:specinfo}. 
The reference stars adopted in the photometric analysis are listed in Table~\ref{table:reference_star}. 
Additionally, the properties of all comparison SNe II used in this work are summarized in Table~\ref{tab:SNe_II info}.

\begin{table*}
\setlength{\tabcolsep}{6pt}
\centering
\small
\caption{Photometric observations of SN\,2024abfl based on LJT.}
\label{table:photdata}
\begin{tabular}{rrrrrrrrrrrrrrr}
\toprule
MJD & $u$ & $\sigma_u$ & $B$ & $\sigma_B$ & $g$ & $\sigma_g$ & $V$ & $\sigma_V$ & $r$ & $\sigma_r$ & $i$ & $\sigma_i$ & $z$ & $\sigma_z$ \\
\hline
60630.7 & - & - & 17.11 & 0.03 & 17.05 & 0.02 & 16.93 & 0.02 & 17.02 & 0.02 & 17.16 & 0.01 & - & - \\
60631.8 & 17.16 & 0.06 & 17.20 & 0.03 & 17.00 & 0.03 & 16.92 & 0.03 & 16.97 & 0.02 & 17.09 & 0.03 & 17.16 & 0.03 \\
60633.6 & 17.21 & 0.06 & 17.20 & 0.02 & 17.05 & 0.01 & 16.95 & 0.01 & 16.96 & 0.01 & 16.99 & 0.01 & 17.08 & 0.02 \\
60635.7 & 17.36 & 0.07 & 17.26 & 0.01 & 17.07 & 0.01 & 16.94 & 0.01 & 16.93 & 0.01 & 17.01 & 0.01 & 16.92 & 0.02 \\
60638.7 & 17.64 & 0.05 & - & - & - & - & 16.91 & 0.01 & 16.83 & 0.01 & 16.90 & 0.02 & 16.97 & 0.02 \\
60642.7 & - & - & 17.56 & 0.01 & 17.24 & 0.01 & 16.97 & 0.01 & 16.90 & 0.01 & 16.91 & 0.01 & - & - \\
60644.7 & 18.76 & 0.07 & - & - & 17.32 & 0.01 & - & - & 16.89 & 0.01 & 16.90 & 0.01 & 16.89 & 0.02 \\
60648.7 & 19.53 & 0.09 & - & - & 17.43 & 0.01 & - & - & 16.96 & 0.01 & 16.93 & 0.01 & 16.88 & 0.02 \\
60650.7 & - & - & - & - & 17.47 & 0.02 & - & - & 16.90 & 0.01 & 16.89 & 0.01 & 16.65 & 0.02 \\
60653.7 & - & - & - & - & 17.49 & 0.03 & - & - & 16.84 & 0.03 & 16.82 & 0.02 & 16.57 & 0.02 \\
60656.8 & - & - & - & - & 17.53 & 0.04 & - & - & 16.91 & 0.02 & 16.89 & 0.02 & - & - \\
60657.7 & - & - & - & - & 17.66 & 0.04 & - & - & 16.96 & 0.02 & 16.87 & 0.01 & 16.73 & 0.02 \\
60661.5 & 20.19 & 0.21 & - & - & 17.64 & 0.03 & - & - & 16.90 & 0.01 & 16.78 & 0.01 & 16.69 & 0.02 \\
60662.8 & 20.24 & 0.22  & - & - & 17.63 & 0.03 & - & - & 16.91 & 0.01 & 16.79 & 0.01 & 16.62 & 0.01 \\
60668.7 & - & - & - & - & 17.64 & 0.02 & - & - & 16.88 & 0.01 & 16.72 & 0.01 & 16.99 & 0.03 \\
60671.7 & - & - & - & - & 17.71 & 0.01 & - & - & - & - & 16.73 & 0.01 & 16.65 & 0.01 \\
60672.7 & - & - & - & - & 17.69 & 0.01 & - & - & 16.88 & 0.01 & 16.70 & 0.01 & 16.53 & 0.01 \\
60675.6 & - & - & - & - & 17.72 & 0.01 & - & - & 16.85 & 0.01 & 16.68 & 0.01 & 16.56 & 0.08 \\
60688.7 & - & - & - & - & - & - & - & - & - & - & 16.66 & 0.02 & - & - \\
60690.6 & - & - & - & - & 17.80 & 0.03 & - & - & 16.83 & 0.01 & 16.61 & 0.01 & - & - \\
60698.7 & - & - & - & - & 17.84 & 0.01 & 17.19 & 0.01 & 16.82 & 0.01 & 16.58 & 0.01 & 16.39 & 0.01 \\
60704.7 & - & - & - & - & 17.80 & 0.02 & 17.16 & 0.02 & 16.81 & 0.01 & 16.56 & 0.01 & 16.39 & 0.01 \\
60707.7 & - & - & - & - & 17.82 & 0.02 & 17.14 & 0.02 & 16.79 & 0.01 & 16.56 & 0.01 & 16.46 & 0.01 \\
60716.8 & - & - & - & - & - & - & 17.13 & 0.05 & 16.81 & 0.02 & - & - & - & - \\
60721.7 & - & - & - & - & 17.95 & 0.04 & 17.21 & 0.03 & 16.84 & 0.01 & 16.61 & 0.01 & 16.48 & 0.01 \\
60727.6 & - & - & - & - & 17.95 & 0.01 & 17.24 & 0.01 & 16.88 & 0.01 & 16.62 & 0.02 & - & - \\
60733.5 & - & - & - & - & 17.96 & 0.01 & 17.28 & 0.01 & - & - & 16.65 & 0.01 & - & - \\
60739.7 & - & - & - & - & 17.99 & 0.02 & 17.32 & 0.02 & 16.91 & 0.01 & - & - & - & - \\
60744.6 & - & - & - & - & 17.85 & 0.07 & 17.23 & 0.05 & 16.89 & 0.03 & 16.69 & 0.02 & 16.54 & 0.01 \\
60749.6 & - & - & - & - & 18.03 & 0.05 & 17.32 & 0.03 & 16.98 & 0.01 & 16.76 & 0.01 & 16.64 & 0.01 \\
60757.5 & - & - & - & - & 20.61 & 0.20 & 19.73 & 0.14 & 19.01 & 0.05 & 18.49 & 0.04 & - & - \\
60758.5 & - & - & - & - & 21.04 & 0.17 & 19.19 & 0.44 & - & - & - & - & - & - \\
60760.6 & - & - & - & - & - & - & 19.61 & 0.07 & - & - & - & - & - & - \\
60763.6 & - & - & - & - & - & - & - & - & 18.94 & 0.07 & 18.46 & 0.05 & - & - \\
60764.5 & - & - & - & - & - & - & - & - & 19.06 & 0.04 &  18.60 & 0.04 & - & - \\
60767.5 & - & - & - & - & - & - & - & - & 19.01 & 0.08 & 18.52 & 0.04 & - & - \\
60775.5 & - & - & - & - & - & - & - & - & 19.30 & 0.11 & 18.50 & 0.08 & - & - \\
\hline
\end{tabular}
\end{table*}

\begin{table*}[t]
\caption{Log of spectroscopic observations of SN\,2024abfl based on LJT and XLT.}
\label{table:specinfo}
\centering
\small                                      
\setlength{\tabcolsep}{6pt}
\begin{tabular}{c c c c c c c} 
\hline
Date & MJD & Phase$^*$ & Telescope+Instrument & Grism+Slit & Airmass & Exp. time \\ 
\hline 
20241116 & 60630.7  & +7.5d  & LJT+YFOSC & G3+1.81" & 1.83 & 1800 \\
20241117 & 60631.7  & +8.6d  & LJT+YFOSC & G3+1.81" & 1.66 & 2000 \\
20241117 & 60631.9  & +8.7d  & XLT+BYOSC & G4+1.8"  & 1.34 & 3300 \\  
20241119 & 60633.7  & +10.5d  & LJT+YFOSC & G3+1.81" & 1.83 & 2200 \\
20241119 & 60633.9  & +10.7d  & XLT+BYOSC & G4+1.8"  & 1.35 & 3300 \\ 
20241128 & 60642.8  & +19.6d  & LJT+YFOSC & G3+1.81" & 1.62 & 2000 \\
20241204 & 60648.7  & +25.6d  & LJT+YFOSC & G3+1.81" & 1.62 & 1500 \\
20241206 & 60650.1  & +27.5d  & LJT+YFOSC & G3+1.81" & 1.64 & 1500 \\
20241209 & 60653.6  & +30.5d  & LJT+YFOSC & G3+2.51" & 1.77 & 2000 \\
20241213 & 60657.7  & +34.5d  & LJT+YFOSC & G3+2.51" & 1.63 & 2400 \\
20241218 & 60662.9  & +39.7d  & LJT+YFOSC & G3+2.51" & 1.76 & 2200 \\
20241222 & 60666.9  & +43.8d  & XLT+BYOSC & G4+1.8"  & 1.57 & 3300 \\ 
20241224 & 60668.7  & +45.6d  & LJT+YFOSC & G3+2.51" & 1.61 & 1800 \\
20241228 & 60672.7  & +49.5d  & LJT+YFOSC & G3+2.51" & 1.61 & 2400 \\
20250105 & 60680.7  & +57.6d  & LJT+YFOSC & G3+2.51" & 1.62 & 2300 \\
20250110 & 60685.7  & +62.5d  & LJT+YFOSC & G3+2.51" & 1.61 & 2200 \\
20250113 & 60688.7  & +65.5d  & LJT+YFOSC & G3+2.51" & 1.61 & 2200 \\
20250120 & 60695.7  & +72.5d  & LJT+YFOSC & G3+2.51" & 1.61 & 2300 \\
20250201 & 60707.7  & +84.5d  & LJT+YFOSC & G3+2.51" & 1.70 & 2400 \\
20250210 & 60716.8  & +93.6d  & LJT+YFOSC & G3+2.51" & 1.97 & 2600 \\
20250222 & 60728.7  & +105.5d  & LJT+YFOSC & G3+1.81" & 1.76 & 2400 \\
20250305 & 60739.7  & +116.5d  & LJT+YFOSC & G3+1.81" & 1.94 & 2500 \\
20250326 & 60760.5  & +137.4d  & LJT+YFOSC & G10+2.51" & 1.65 & 2500 \\
\hline
\multicolumn{7}{l}{{$^*$Phases are relative to the estimated explosion epoch (MJD = $57837.1\, \pm 0.1$) in observer frame.}}
\end{tabular}
\end{table*}

\begin{table*}
\centering
\small
\setlength{\tabcolsep}{3pt}  
\caption{Reference stars in the SN\,2024abfl field. Most of the photometric error from VizieR is below 0.01\,mag.}
\label{table:reference_star}
\begin{tabular}{cccccccccc}
\hline
Number & RA & Dec & $B$ & $V$ & $u^*$ & $g$ & $r$ & $i$ & $z$ \\
\hline
1 & $\rm 06^h17^m31.76^s$ & $+78^{\circ}25^{\prime}28.18^{\prime\prime}$& 14.27(0.06)&13.32(0.00)&-&13.76(0.00)&13.21(0.00)&12.91(0.00)&12.74(0.00)\\
2 & $\rm 06^h16^m59.64^s$ & $+78^{\circ}24^{\prime}20.26^{\prime\prime}$& 14.59(0.00)&13.79(0.00)&15.53(0.03)&14.15(0.00)&13.60(0.01)&13.45(0.00)&13.34(0.00)\\
3 & $\rm 06^h17^m14.35^s$ & $+78^{\circ}23^{\prime}07.51^{\prime\prime}$& 14.86(0.00)&13.95(0.00)&16.17(0.03)&14.39(0.00)&13.76(0.00)&13.52(0.00)&13.44(0.00)\\
4 & $\rm 06^h17^m44.11^s$ & $+78^{\circ}21^{\prime}17.85^{\prime\prime}$& 15.39(0.00)&14.77(0.00)&16.11(0.05)&15.19(0.04)&14.65(0.00)&14.50(0.00)&14.48(0.00)\\
5 & $\rm 06^h17^m52.22^s$ & $+78^{\circ}19^{\prime}47.27^{\prime\prime}$& 14.36(0.06)&13.69(0.02)&15.28(0.08)&14.01(0.00)&13.57(0.02)&13.44(0.00)&13.44(0.00)\\
6 & $\rm 06^h19^m16.47^s$ & $+78^{\circ}20^{\prime}40.03^{\prime\prime}$& 15.21(0.00)&14.50(0.00)&-&14.94(0.00)&14.46(0.00)&14.28(0.01)&14.22(0.00)	\\
\hline
\end{tabular}
\begin{flushleft}
*The multi-band photometric data of the reference stars were retrieved from VizieR, except for the $u/U$ band. 
The $u$-band magnitudes here were estimated using transformation equations of \citet{2005AJ....130..873J}, while the supplementary $U$-band data were obtained from \citet{2020MNRAS.498...84Z}, where the photometry was calibrated to the absolute fluxes of standard stars.
\end{flushleft}
\end{table*}

\begin{table*}
\centering
\small
\setlength{\tabcolsep}{5pt}  
\caption{Quantities that were derived and used in EPM.}
\label{table:EPM quantities}
\begin{tabular}{cccccccccc}
\hline
MJD & Phase(d) & $T(\mathrm{K})$ & $v_{\rm ph}$ & $\xi_{\rm E96}$ & $\theta_{\rm E96}\,(10^{-12}\,\mathrm{rad})$ & $\xi_{\rm D05}$ & $\theta_{\rm D05}\,(10^{-12}\,\mathrm{rad})$ & $\xi_{\rm V19}$ & $\theta_{\rm V19}\,(10^{-12}\,\mathrm{rad})$\\
\hline
60630.7 & 7.5 & 9658(231) & 5530(989) & 0.396 & 8.316 & 0.559 & 5.888 & 0.470 & 7.005 \\
60631.8 & 8.6 & 8938(155) & 4042(753) & 0.416 & 8.828 & 0.596 & 6.157 & 0.502 & 7.310 \\
60633.7 & 10.5 & 8276(70) & 3868(726) & 0.444 & 9.396 & 0.641 & 6.510 & 0.546 & 7.648 \\
\hline
\end{tabular}
\begin{flushleft}
\end{flushleft}
\end{table*}

\begin{table*}[t]
    \centering
    \small
    \setlength{\tabcolsep}{14pt} 
    \caption{Properties of the comparison SNe II.}
    \begin{tabular}{cccccc}
        \hline
        SN & Explosion Date & Redshift & Distance & $E(B-V)_\mathrm{tol}$ & References \\
         & [MJD] & z & [Mpc] & [mag] & \\
        \hline
        1999em & 51475.1$\pm$1.4 & 0.00239 & 8.2$\pm$2.6 & 0.10$\pm$0.05 & 1 \\
        2005cs & 53548.5$\pm$1.0 & 0.00153 & 7.1$\pm$1.2 & 0.05 & 2 \\
        2009N &  54847.6 & 0.0035 & 21.6$\pm$1.1 & 0.13$\pm$002 & 3 \\
        2008in & 54825.1 & 0.005224 & 13.19$\pm$1.09 & 0.0984$\pm$0.104 & 4 \\
        2016bkv & 57467.5$\pm$1.2 & 0.001975 & 14.4$\pm$0.3 & 0.015 & 5\\
        2017eaw & 57886.5$\pm$0.1 & 0.000133 & 6.85$\pm$0.30 & 0.41 & 6 \\
        2018is & 58132.9$\pm$1.1 & 0.005811 & 21.3$\pm$1.7 & 0.19$\pm$0.06 & 7 \\
        2018zd & $58178.39^{+0.15}_{-0.50}$ & 0.002979 & $15.0^{+8.9}_{-1.9}$ & 0.17$\pm$0.05 & 8 \\
        2018lab & 58480.4$\pm$0.1 & 0.0089 & 35.5 & 0.22 & 9 \\
        2020cxd & 58897.0$\pm$1.5 & 0.003883 &  22.0$\pm$3.0 & 0.035 & 10\\
        2021yja & 59464.4$\pm$0.1& 0.005307 & $23.4^{+5.4}_{-4.4}$ & 0.104 & 11 \\
        2022acko & 59918.17 & 0.00526 &19.8$\pm$2.8 & 0.056$\pm$0.01& 12 \\
        2024abfl &  $60623.16^{+0.14}_{-0.07}$ & 0.002979 & -- & 0.19 & This work\\
        \hline
    \end{tabular}%
    \begin{flushleft}
    \small
    References: 
    1. \citet{2002PASP..114...35L,2003MNRAS.338..939E},
    2. \citet{2006MNRAS.370.1752P,2009MNRAS.394.2266P},
    3. \citet{2014MNRAS.438..368T},
    4. \citet{2011ApJ...736...76R},
    5. \citet{2018ApJ...859...78N,2018ApJ...861...63H},
    6. \citet{2019ApJ...876...19S,2019ApJ...875..136V},
    7. \citet{2025A&A...694A.260D},
    8. \citet{2020MNRAS.498...84Z},
    9. \citet{2023ApJ...945..107P},
    10. \citet{2021A&A...655A..90Y,2022MNRAS.513.4983V}
    11. \citet{2022ApJ...935...31H},
    12. \citet{2025MNRAS.540.2591L,2023ApJ...953L..18B},
    \end{flushleft}
    \label{tab:SNe_II info}
\end{table*}

\end{document}